\documentclass[twocolumn,showpacs,showkeys,superscriptaddress,nofootinbib,floatfix]{revtex4-1}
\usepackage{amsmath}
\usepackage{graphicx}
\usepackage{bm}

\makeatletter
\newcommand\erfc{\mathop{\operator@font erfc}\nolimits}
\def\slashchar#1{\setbox0=\hbox{$#1$}
   \dimen0=\wd0 \setbox1=\hbox{/} \dimen1=\wd1
   \ifdim\dimen0>\dimen1 \rlap{\hbox to \dimen0{\hfil/\hfil}} #1
   \else  \rlap{\hbox to \dimen1{\hfil$#1$\hfil}} / \fi}

\makeatother

\begin{document}

\title{Generalized Quark Transversity Distribution of the Pion in Chiral
Quark Models}
\author{Alexander E. Dorokhov}
\affiliation{Joint Institute for Nuclear Research, Bogoliubov Laboratory of Theoretical
Physics, 114980, Dubna, Russia}
\affiliation{Institute for Theoretical Problems of Microphysics, Moscow State University,
RU-119899, Moscow, Russia}
\email{dorokhov@theor.jinr.ru}
\author{Wojciech Broniowski}
\affiliation{The H. Niewodnicza\'nski Institute of Nuclear Physics, Polish Academy of
Sciences, PL-31342 Krak\'ow, Poland}
\affiliation{Institute of Physics, Jan
Kochanowski University, PL-25406~Kielce, Poland}
\email{Wojciech.Broniowski@ifj.edu.pl}
\author{Enrique Ruiz Arriola}
\affiliation{Departamento de F\'{\i}sica At\'omica, Molecular y Nuclear, Universidad de
Granada, E-18071 Granada, Spain}
\affiliation{Instituto Carlos I de Fisica Te\'orica y Computacional, Universidad de Granada, E-18071 Granada, Spain}
\email{earriola@ugr.es}
\date{\today}

\begin{abstract}
The transversity generalized parton distributions (tGPDs) of the the
pion, involving matrix elements of the tensor bilocal quark current,
are analyzed in chiral quark models. We apply the nonlocal chiral
models involving a momentum-dependent quark mass, as well as the local
Nambu--Jona-Lasinio with the Pauli-Villars regularization to
calculate the pion tGPDs, as well as related quantities following
from restrained kinematics, evaluation of moments, or taking the
Fourier-Bessel transforms to the impact-parameter space.  The obtained
distributions satisfy the formal requirements, such as proper support
and polynomiality, following from Lorentz covariance. We carry out the
leading-order QCD evolution from the low quark-model scale to higher
lattice scales, applying the method of Kivel and Mankiewicz. We
evaluate several lowest-order generalized transversity form factors,
accessible from the recent lattice QCD calculations. These form factors,
after evolution, agree properly with the lattice data, in support of the
fact that the spontaneously broken chiral symmetry is the key element
also in the evaluation of the transversity observables.
\end{abstract}

\maketitle

\section{Introduction \label{sec:intro}}

The underlying spin-$\frac{1}{2}$ partonic structure of hadrons became
first manifest in the analysis of the deep inelastic
scattering~\cite{Callan:1969uq}. Actually, further understanding of
the partonic spin distributions can be gained by the study of the {\em
  transversity distributions}~\cite{Barone:2001sp}. From this
viewpoint, generalized parton distributions (GPDs) \cite%
{Mueller:1998fv,Ji:1996ek,Radyushkin:1996nd} (for extensive reviews
see, e.g.,~\cite{Belitsky:2005qn,Feldmann:2007zz,Boffi:2007yc} and
references therein) encode a detailed information on the parton
structure of hadrons when analyzed at short distances. In the
impact-parameter space, the GPD's can be viewed as partonic
probabilities in the infinite-momentum frame distributed along the
longitudinal momentum fraction (Bjorken-x) and the transverse space
directions~\cite{Burkardt:2000za,Burkardt:2002hr}. It should be noted
that both GPD's as well as their partonic interpretation depend
strongly on the renormalization scale and it is not obvious {\it a
  priori} what, if any, is the reference scale, which might have some
universal value and significance. From a dynamical point of view, the
choice of such a scale is crucial, as the high-energy modes are
integrated out in favor of an effective and yet unknown
non-perturbative low-energy dynamics.  The renormalization group deals
with the intertwining of scales in principle, although in practice it
can be explored only at the lowest orders of the perturbation theory
in the running strong coupling constant. In addition, GPD's depend
also on the factorization scheme corresponding to the physical process
used to extract the partonic distributions at high energies.

>From a purely theoretical point of view, the great difficulty to
determine the GPDs from first principles in QCD is related to their
genuine Minkowski-space nature, suggesting application of the light-cone
kinematics and non-perturbatively motivated approaches, such as the
transverse lattice~\cite{Burkardt:2001jg}, which so far has produced
encouraging but scarce results. More recently, however, the lowest
Bjorken-$x$ moments of the kinematically intricate GPDs, the so-called
Generalized Form Factors (GFFs), have become directly accessible to
Euclidean lattices in QCD at sufficiently short-distance resolution
scales (see, e.g., \cite{Musch:2010ka,Hagler:2009ni}). This is due to
the fact that GFFs for space-like momenta can be written as matrix
elements of local operators which can be directly extracted from the
asymptotics of the Euclidean correlation functions. As a further
simplification, the scale dependence of GFFs in the space-like region
undergoes a triangular-matrix multiplicative renormalization, which can
be easily implemented (see, e.g.,~\cite{Broniowski:2009zh}). A well
known feature of the QCD evolution is the loss of resolution at higher
energies, a property triggered by the existence of the asymptotic
ultraviolet fixed point, which enhances similarity at increasingly
high $Q^2$-values.

In this paper we analyze the quark \emph{transversity} generalized
parton distribution of the pion (tGPD), related to the matrix elements
of the bilocal tensor current operator $\bar{q}(x)\sigma _{\mu \nu
}q(0)$ (see Sec.~\ref{sec:def} and Refs.~\cite{Diehl:2005jf,Burkardt:2005hp} for
precise definitions). The transversity distribution, also termed the
\emph{maximal helicity} GPD, as it involves aligned parton-helicity
operators, provides insight into the nontrivial spin structure of the
hadron. For the spin-0 hadrons, tGPDs arise due to a nonzero orbital angular
momentum between the initial and final state, and thus offer a unique
opportunity to learn about the spin structure without the many
complications of the hadronic spin degrees of freedom, as is the case of the nucleon.  Due to their
inherent complexity, tGPDs are the least investigated among the hadronic GPD's.
In this regard the study of the spin structure of the pion is
particularly appealing and challenging, although at present it is
unclear how it can be reliably extracted from the high-energy
experiments.

The recent lattice determination of the first two $X$-moments
of the pion tGPD, denoted as \emph{transversity} generalized form factors (tGFFs)~\cite{Brommel:2007xd}, provides
first important and non-trivial information on this
issue. The calculation was carried out at a lattice spacing of $a \sim 0.1 {\rm ~fm}$ and a pion mass $m_\pi \sim 600 {\rm ~MeV}$. For such a
small lattice spacing the matching to the perturbative $\overline{\rm MS}$ scheme becomes feasible and corresponds to
the scale $\mu \simeq 2 {\rm~GeV}$. This lattice calculation has triggered some related studies
focusing either on perturbative aspects of the high-$Q^{2}$ dependence
of the transversity form factors~\cite{Diehl:2010ru}, or non-perturbative issues studied within chiral quark
models~\cite{Broniowski:2010nt,Nam:2010pt}.

In this work we analyze the tGPD and the tGFFs of the pion for several
chiral quark models, extending the results presented previously~\cite{Broniowski:2010nt} and
providing further details. While this
unavoidably makes the paper a bit technical, we hope that many of the
details provided here show how a proper implementation of the chiral symmetry,
relativity, and normalization can be achieved in a non-perturbative model
calculation. This is particularly
interesting for the case of nonlocal models, where the mass function
depends on the momentum.  Although such models are expected to feature chiral
quark dynamics more realistically, many complications arise due to the
time-like kinematics implied by the very definition of the GPDs. We recall
that we are effectively carrying out the one-loop calculations, where some
variables are integrated out and some may be left unintegrated. Thus, special attention
must be paid to the treatment of the integrals, particularly to keep
the Poincar\'e invariance explicitly at any step of the calculation,
such that all results are mutually consistent.

Via sum rules, the (generalized) form factors are related to the GPDs~\cite%
{Ji:1998pc,Radyushkin:2000uy,Goeke:2001tz,Bakulev:2000eb,Diehl:2003ny,Ji:2004gf,Belitsky:2005qn,Feldmann:2007zz,Boffi:2007yc}%
. Experimentally, the GPDs of the pion constitute rather elusive
quantities which appear in rare exclusive processes, such as the
deeply virtual Compton scattering (DVCS) or the hard electro-production of
mesons (HMP).

Chiral quark models have proved to correctly describe numerous
features related to the vector GPD of pion. The parton distribution
functions (PDF) have been evaluated in the Nambu--Jona-Lasinio (NJL)
model in Refs.~\cite%
{Davidson:1994uv,RuizArriola:2001rr,Davidson:2001cc}. The extension to
diagonal GPDs in the impact parameter space was carried out in \cite%
{Broniowski:2003rp}. Other analyses of the pionic GPDs and PDFs were
performed in nonlocal chiral quark models \cite%
{Dorokhov:1998up,Polyakov:1999gs,Dorokhov:2000gu,Anikin:2000th,Praszalowicz:2002ct,Praszalowicz:2003pr,Bzdak:2003qe,Holt:2010vj,Nguyen:2011jy}%
, in the NJL model \cite%
{Polyakov:1999gs,Theussl:2002xp,Bissey:2003yr,Noguera:2005cc,Broniowski:2007si}
and in the light-front constituent quark models~\cite%
{Frederico:2009pj,Frederico:2009fk}. The parton distribution
amplitudes, related to the GPD via a low-energy
theorem~\cite{Polyakov:1998ze}, were evaluated in~\cite%
{Esaibegian:1989uj,Dorokhov:1991nj,Petrov:1998kg,Anikin:1999cx,Praszalowicz:2001wy,Dorokhov:2002iu,RuizArriola:2002bp,RuizArriola:2002wr}%
. The gravitational form factors were computed in
\cite{Broniowski:2008hx}.  Finally, the pion-photon transition
distribution amplitudes \cite%
{Pire:2004ie,Pire:2005ax,Lansberg:2006fv,Lansberg:2007bu} were
obtained in Refs.~\cite%
{Tiburzi:2005nj,Broniowski:2007fs,Courtoy:2007vy,Courtoy:2008af,Kotko:2008gy}%
.

Besides the phenomenological motivation, it is useful to review
shortly what aspects of the present investigation suggest the use of
chiral quark models within the present context (see, e.g.,
\cite{RuizArriola:2002wr}). Firstly, the pion, treated as a composite $q \bar
q$ state, becomes a Goldstone boson of the spontaneously broken chiral
symmetry. This of course requires the correct implementation of the
chiral Ward-Takahashi identities -- a rather non-trivial point, since
this condition is not automatically fulfilled in loop
calculations. At the quark level, this feature is compatible with the
large-$N_c$ scaling relations. Within such a scheme the pion loop
corrections are $1/N_c$-suppressed but chiral-log enhanced at small
pion masses. However, the leading-$N_c$ contributions present a much
milder pion-mass dependence, a favorable situation for the
unphysically large pion masses used on the
lattice~\cite{Brommel:2007xd}. Moreover, relativity for the GPDs is
properly implemented through the so-called polynomiality conditions,
and, more specifically, by the explicit use of the double
distributions (DDs). Finally, the scale at which a quark model
calculation is carried out can only be identified after a correct
separation of the momentum fraction carried by the quark degrees of
freedom.  As mentioned already, the partonic properties depend on the
renormalization scale, and according to
phenomenology~\cite{Sutton:1991ay,Gluck:1999xe} as well as independent
lattice calculations~\cite{Best:1997qp}, the (valence) quarks carry about
40\% of the total momentum at the scale $\mu= 2{\rm GeV}$. In
effective quark models, where the quarks carry 100\% of the total
momentum, the perturbative scale is unexpectedly and rather
uncomfortably low. However, the assumption has been tested to higher
orders and confronted by comparing to a variety of
high-energy data or lattice calculations. In the present calculation
of the transversity form factors we find again agreement with the data
after the QCD evolution scheme is implemented, starting from a low
quark-model scale.

GPDs in general, and tGPDs in particular, are subjected to a set of conditions {\it
a priori} imposed by symmetries and/or completeness, namely, the chiral
symmetry, relativity, positivity, and finiteness of sum rules.  Within
the framework of low energy chiral quark models, where there is an
inherent cut-off marking the low energy regime, these conditions are
actually not easy to fulfill on purely mathematical grounds.  Indeed,
one-loop integrals are four dimensional, whereas GPDs leave two
integration variables unintegrated and hence some consistency is
required. However, once this difficulty is mastered, which is the case
of our approach, there is a trend to
independence to details of the model. This independence is largely enhanced
{\it after} the QCD evolution, since differences are washed out at
increasingly higher energy scales.  The feature is also observed
in the study of transversity, as to make differences between
various chiral quark models rather small.

We apply the local NJL model with the Pauli-Villars regularization, as well as two
variants of the nonlocal chiral quark models inspired by the
nontrivial structure of the QCD vacuum~\cite%
{Diakonov:1985eg,Holdom:1990iq}. These models provide the results at
the quark-model scale. After the necessary (multiplicative) QCD
evolution~\cite{Broniowski:2007si}, our model results are in a quite
remarkable agreement with the lattice data for tGFFs. Lower values of
the constituent quark mass, $\sim 250$~MeV, are preferred.

The outline of the paper is as follows: In Sec.~\ref{sec:basics} we
give the general definitions of the pion tGPD and tGFFs. Then we
derive these quantities in the nonlocal chiral quark models from the
triangle diagram in Sec.~\ref{sec:mod}. By using the extremely convenient
$\alpha $-representation, we obtain the corresponding expressions for the tGFFs
in the momentum- and impact-parameter spaces, the tGPDs for the isosinglet and
isovector channels, and also, in special forward and symmetric
kinematics, the distribution of the transversity size of the pion. The analysis is carried out
explicitly for specific nonlocal models in Sec.~\ref{sec:specific}.
For numerical estimates of these quantities we use two variants of the
chiral nonlocal models and the local NJL model. In Sec.~\ref{sec:evol}
we present the QCD evolution of the above quantities in general, as well as show
its consequences for the studied models. Numerical results for
the transversity distribution functions after evolution are shown in
Sec.~\ref{sec:res}. Finally, in Sec.~\ref{sec:concl} we draw our
main conclusions.

\section{Basic definitions of the transversity form factors and generalized parton distribution \label{sec:def}\label{sec:basics}}

In this section we provide the basic definitions as well as the
kinematics of the transversity observables analyzed in the present
work.

The pion $u$-quark tGFFs, $B_{Tni}^{\pi ,u}\left( t\right) $, parametrize the
matrix element
\begin{align}
& \left\langle \pi ^{+}\left( p^{\prime }\right) \left\vert O_{T}^{\mu \nu
\mu _{1}\cdots \mu _{n-1}}\right\vert \pi ^{+}\left( p\right) \right\rangle =%
\mathcal{TAS}\frac{P^{\mu }q^{\nu }}{m_{\pi }}
 \nonumber \\
& \times \sum_{\substack{ i=0,  \\ \mathrm{even}}}^{n-1}q^{\mu _{1}}...q^{\mu
_{i}}P^{\mu _{i+1}}...P^{\mu _{n-1}}B_{Tni}^{\pi ,u}\left( t\right) ,
\label{PionME}
\end{align}%
where the local tensor quark operator is%
\begin{align}
& \mathcal{O}_{T}^{\mu \nu \mu _{1}\cdots \mu _{n-1}}  \label{TensorOp} \\
& =\mathcal{T}\underset{\left( \mu \nu \right) }{\mathcal{A}}\underset{%
\left( \mu _{1}\cdots \mu _{n-1}\right) }{\mathcal{S}}\overline{u}\left(
0\right) i\sigma ^{\mu \nu }i\overleftrightarrow{D}^{\mu _{1}}\cdot \cdot
\cdot i\overleftrightarrow{D}^{\mu _{n-1}}u\left( 0\right) ,  \notag
\end{align}%
with $\overleftrightarrow{D}^{\beta }=\overleftrightarrow{\partial }^{\beta
}-igA^{\beta }$ being the QCD covariant derivative, and $\overleftrightarrow{%
\partial }^{\beta }=\frac{1}{2}\left( \overrightarrow{\partial }^{\beta }-%
\overleftarrow{\partial }^{\beta }\right) $. In Eq.~(\ref{PionME}), $p^{\prime }$
and $p$ are the initial and final pion momenta, while $P=\frac{1}{2}(p^{\prime
}+p) $, $q=p^{\prime }-p$, and $t=-q^{2}.$ The symbol $\mathcal{TAS}$
denotes symmetrization ($\mathcal{S}$) in $\nu ,\mu _{1},\ldots ,\mu _{n-1}$%
, followed by antisymmetrization ($\mathcal{A}$) in $\mu ,\nu $, with the
additional prescription that the traces in all index pairs are subtracted ($%
\mathcal{T}$). The factor $1/m_{\pi }$ is introduced by convention in order
to have dimensionless form factors~\cite{Brommel:2007xd}. Also, as in~\cite%
{Brommel:2007xd}, we use the positively charged pion and the
up-quark density for definiteness.

The above definition, which projects on twist-2 operators, can be
implemented in a simple and manifestly covariant way (see, e.g., \cite{Diehl:2010ru}) by a contraction
with two constant auxiliary four-vectors, $a$ and $b$, satisfying $a^{2}=(ab)=0$ and $%
b^{2}\neq 0$. The tGFFs are then defined via%
\begin{align}
& M_{Tn}^{\pi ,u}\left( \xi ,t\right)  \label{PionTme} \\
& =\left\langle \pi ^{+}\left( p^{\prime }\right) \left\vert \overline{u}%
\left( 0\right) i\sigma ^{\mu \nu }a_{\mu }b_{\nu }\left( i%
\overleftrightarrow{D}a\right) ^{n-1}u\left( 0\right) \right\vert \pi
^{+}\left( p\right) \right\rangle  \notag \\
& =\left( aP\right) ^{n-1}\frac{\left[ \left( ap\right) \left( bp^{\prime
}\right) \right] }{m_{\pi }}\sum_{\substack{ i=0,  \\ \mathrm{even}}}%
^{n-1}\left( 2\xi \right) ^{i}B_{Tni}^{\pi ,u}\left( t\right) ,  \notag
\end{align}%
where the skewness parameter is defined as\footnote{Throughout this work we use the so-called symmetric notation.}
\begin{equation}
\xi =-\frac{\left( aq\right) }{2\left( aP\right) },  \label{PqKsi}
\end{equation}%
$\xi \in [0,1]$, and $(aq)$, etc., denote the scalar products of four-vectors.
In Eq.~ (\ref{PionTme}), $\left[ ...\right] $ denotes the antisymmetrization in $a$
and $b$.

The tGFFs defined in (\ref{PionTme}) refer to the $u$-quarks; those for the $d$%
-quarks follow from the isospin symmetry and read%
\begin{equation}
B_{Tni}^{\pi ,d}\left( t\right) =\left( -1\right) ^{n}B_{Tni}^{\pi ,u}\left(
t\right) .  \label{PionTmeD}
\end{equation}

The definition of the corresponding tGPD is \cite{Belitsky:2005qn}
\begin{eqnarray}
&& \langle \pi ^{+}(p^{\prime })\mid \bar{u}(-a)i\sigma ^{\mu \nu }a_{\mu}b_{\nu }u(a)\mid \pi ^{+}(p)\rangle   \nonumber \\
&& =\frac{\left[ \left( ap\right) \left( bp^{\prime }\right) \right] }
{m_{\pi }}\int_{-1}^{1}dX \,e^{-i X\left( Pa\right) }E_{T}^{\pi, u}(X,\xi ,t), \label{PionTGPD}
\end{eqnarray}
where we do not display explicitly the gauge link factor.
The tGFFs can be written as the Mellin moments of tGPD of the pion as%
\begin{equation}
\int_{-1}^{1}dX\,X^{n-1}E_{T}^{\pi, u}\left( X,\xi ,t\right)
=\sum_{\substack{ i=0, \\ \mathrm{even}}}^{n-1}\left( 2\xi \right)^{i}B_{Tni}^{\pi ,u}\left( t\right).  \label{En}
\end{equation}

\section{Chiral quark models \label{sec:mod}}

In this section we review the generic one-loop features of chiral
quark models, where the quark self-energy as well as the interaction
vertices are assumed to have a fairly general momentum dependence to
be specified later on. We derive general expressions for the tGPD at the
one-quark-loop level, applicable to both nonlocal and local models.
We also display formal properties of tGPD in our aproach.

\subsection{Nonlocal chiral quark models \label{sec:inst}}

In the quark-model calculation in the large-$N_c$ limit the matrix element (\ref{PionTme}) is given by the triangle
diagram shown in Fig.~\ref{fig:tri}\footnote{%
We should emphasize at this point that the tensor matrix element (\ref%
{PionTme}) can not be induced by tadpole-type of diagrams. This is evident,
because these diagrams depend only on one external vector $q$ from which it is
impossible to construct the antisymmetric combination involving the matrix
element (\ref{PionTme}). In this aspect, the results obtained in \cite%
{Nam:2010pt} can not be correct.}. To calculate this diagram we explore
the manifestly covariant method  based on the effective approach to nonperturbative
QCD dynamics. All expressions will be computed in the Euclidean space, appropriate
for the process under consideration and, in general, for the treatment of nonperturbative
physics. The nonperturbative quark propagator, dressed by the interaction
with the QCD vacuum, is assumed to have the form%
\begin{equation}
S\left( k\right) =\frac{\widehat{k}+m\left( k^{2}\right) }{D\left(
k^{2}\right) }.  \label{Qprop}
\end{equation}%
The main requirement imposed on the quark propagator is that at large quark
virtualities one recovers the perturbative limit,
\begin{equation}
S\left( k\right) \overset{k^{2}\rightarrow \infty }{\rightarrow }\frac{%
\widehat{k}}{k^{2}}.  \label{QpropAS}
\end{equation}%
It is also assumed that the dynamical quark mass, $m(k^2)$, is a function rapidly
dropping with the quark virtuality $k^{2}$. It is normalized at zero as%
\begin{equation}
m\left( 0\right) =M_{q},\qquad D\left( 0\right) =M_{q}^{2}.  \label{M0}
\end{equation}%
We also need the quark-pion vertex\footnote{In this work we use the dominant (in the spontaneous symmetry-breaking mechanism)
structures for the quark propagator
and the quark-pion vertex. More general structures are used in the Schwinger-Dyson approach \cite{Maris:1997hd}.}
\begin{equation}
\Gamma _{\pi }^{a}\left( k,q\right) =\frac{i}{f_{\pi }}\gamma _{5}\tau
^{a}F\left( k_{+}^{2},k_{-}^{2}\right) ,  \label{QPiVert}
\end{equation}%
where $k_{\pm }=k\pm q/2$. The nonlocal vertex $F\left(
k_{+}^{2},k_{-}^{2}\right) $ is a symmetric function of its arguments,
normalized to $F\left( k^{2},k^{2}\right) = m\left( k^{2}\right) $. In the
present study, the nonlocal model calculations are performed in the strict chiral
limit, which means that $m\left( k^{2}\rightarrow \infty \right) =0$.

\subsection{Calculation of the triangle diagram \label{sec:triangle}}

\begin{figure}[tb]
\includegraphics[width=.3\textwidth]{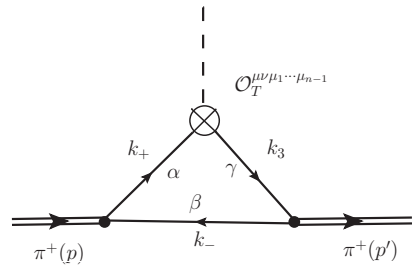} \vspace{-2mm}
\caption{(Color online) The leading-$N_{c}$ one-quark-loop triangle diagram
contribution to the leading twist tGPD of the pion.
\label{fig:tri}}
\end{figure}

Within the described approach the triangle diagram for the matrix element (\ref%
{PionTme}) yields
\begin{align}
& M_{Tn}\left( \xi ,t\right) =\frac{N_{c}}{4\pi ^{2}f_{\pi }^{2}}\int \frac{%
d^{4}k}{\pi ^{2}}F\left( k_{+}^{2},k_{-}^{2}\right) F\left(
k_{3}^{2},k_{-}^{2}\right) \\
& \frac{1}{4}Tr\left\{ S\left( k_{+}\right) \gamma _{5}S\left( k_{-}\right)
\gamma _{5}S\left( k_{3}\right) \sigma _{\mu \nu }\right\} \left( \frac{%
k_{+}+k_{3}}{2},a\right) ^{n-1} \!\!\!\!\! a_{\mu }b_{\nu },  \notag
\end{align}%
where $k_{+}=k$ is the initial momentum of the struck quark, $k_{3}=k_{+}+q$ is its final
momentum, $k_{-}=k_{+}-p$ is the momentum of the spectator quark (cf. Fig.~\ref{fig:tri}), and the covariant
average momentum  $(k_{+}+k_{3})/2$ corresponds to the
derivative in the definition~(\ref{PionTme}).

After taking the trace one has%
\begin{align}
& M_{Tn}\left( \xi ,t\right) =\frac{N_{c}}{4\pi ^{2}f_{\pi }^{2}}\int \frac{%
d^{4}k}{\pi ^{2}}\frac{F\left( k_{+}^{2},k_{-}^{2}\right) F\left(
k_{3}^{2},k_{-}^{2}\right) }{D\left( k_{+}^{2}\right) D\left(
k_{-}^{2}\right) D\left( k_{3}^{2}\right) }  \label{MT} \\
& \times \left\{ m\left( k_{+}^{2}\right) \left[ \left( k_{-}a\right) \left(
k_{3}b\right) \right] -m\left( k_{-}^{2}\right) \left[ \left( k_{+}a\right)
\left( k_{3}b\right) \right] \right.  \notag \\
& \left. +m\left( k_{3}^{2}\right) \left[ \left( k_{+}a\right) \left(
k_{-}b\right) \right] \right\} \left( \frac{k_{+}+k_{3}}{2},a\right) ^{n-1},
\notag
\end{align}%
where the antisymmetrization in $a$ and $b$ is implied. Considering the crossed channel
it is easy to get the relation
\begin{align}
& \left( \left\{ ...\right\} \left( \frac{k_{+}+k_{3}}{2},a\right)
^{n-1}\right) _{d\mathrm{-channel}} \\
& \rightarrow \left( -1\right) ^{n}\left( \left\{ ...\right\} \left( \frac{%
k_{+}+k_{3}}{2},a\right) ^{n-1}\right) _{u\mathrm{-channel}},  \notag
\end{align}%
in agreement with (\ref{PionTmeD}).

For the further analysis, it is very convenient to transform the integral in (\ref{MT}%
) into the $\alpha $-representation (see \cite{Bogolyubov:1980,Zavialov:1990}%
), which is one of the basic methods for the study of hard processes in
perturbative QCD \cite{Radyushkin:1997ki}, as well as in nonperturbative
quark models \cite{Dorokhov:2000gu}. The technical advantage of this method is the
explicit maintenance of the Lorentz covariance.

Let us define for any function $F$ of virtuality $k^{2}$, decaying at large
virtuality as $1/k^{2}$ or faster, its $\alpha $ representation (Laplace
transform)
\begin{equation}
F\left( k^{2}\right) =\int_{0}^{\infty }d\alpha \, e^{-\alpha k^{2}}f\left(
\alpha \right) 
\label{AlphaDef}
\end{equation}%
where $F\left( k^{2}\right) $ is the image of the original function $f\left( \alpha
\right) $. We will use the short-hand $F\left( k^{2}\right) \sim f\left( \alpha
\right)  $. Let us introduce the following notation \cite{Dorokhov:2010bz,Dorokhov:2010zzb}
\begin{align}
& \frac{F(k_{+}^{2},k_{-}^{2})F\left( k_{3}^{2},k_{-}^{2}\right) }{D\left(
k_{+}^{2}\right) D\left( k_{-}^{2}\right) D\left( k_{3}^{2}\right) }m\left(
k_{+}^{2}\right) \sim G_{m,0,0}\left( \alpha ,\beta ,\gamma \right) ,\quad
\notag \\
& \frac{F(k_{+}^{2},k_{-}^{2})F\left( k_{3}^{2},k_{-}^{2}\right) }{D\left(
k_{+}^{2}\right) D\left( k_{-}^{2}\right) D\left( k_{3}^{2}\right) }m\left(
k_{-}^{2}\right) \sim G_{0,m,0}\left( \alpha ,\beta ,\gamma \right) ,  \notag
\\
\quad & \frac{F(k_{+}^{2},k_{-}^{2})F\left( k_{3}^{2},k_{-}^{2}\right) }{%
D\left( k_{+}^{2}\right) D\left( k_{-}^{2}\right) D\left( k_{3}^{2}\right) }%
m\left( k_{3}^{2}\right) \sim G_{0,0,m}\left( \alpha ,\beta ,\gamma \right),
\end{align}%
where the triple $\alpha $ representation (i.e. in parameters $\alpha$,
$\beta$, and $\gamma$) is applied (see Fig. \ref{fig:tri}). With this notation the
momentum integral in Eq.~(\ref{MT}) is transformed into the
$\alpha$-representation expression for the matrix element,
\begin{align}
& M_{Tn}\left( \xi ,t\right) =\left( aP\right) ^{n-1}\left[ \left( ap\right)
\left( bp^{\prime }\right) \right] \frac{N_{c}}{4\pi ^{2}f_{\pi }^{2}}  \label{MTn}
\times \\ & \int \frac{d\left( \alpha \beta \gamma
\right) }{\Delta ^{3}}e^{-\frac{1}{\Delta }\left( \alpha \gamma t-\beta
\left( \alpha +\gamma \right) m_{\pi }^{2}\right) }\left( \frac{\beta
+\left( \gamma -\alpha \right) \xi }{\Delta }\right) ^{n-1}  \notag
\times \\ & \left[ \alpha G_{m,0,0}\left( \alpha ,\beta ,\gamma \right) \! + \!\beta
G_{0,m,0}\left( \alpha ,\beta ,\gamma \right) \! + \! \gamma G_{0,0,m}\left( \alpha
,\beta ,\gamma \right) \right] ,  \notag
\end{align}%
where $\Delta =\alpha +\beta +\gamma $ and
\begin{eqnarray}
\int d\left( \alpha \beta \gamma
\right) ...=\int_{0}^{\infty }d\alpha \int_{0}^{\infty }d\beta
\int_{0}^{\infty }d\gamma ...
\end{eqnarray}

The only dependence on $\xi$ in Eq.~(\ref{MTn}) appears in the polynomial factor in the second line.
It is clear that in the expansion of this polynomial in
powers of $\xi $ only the even powers survive, in accordance with Eq.~(\ref{PionTme}%
), since for the odd powers of $\xi $ the integrand is antisymmetric in $\alpha $
and $\gamma $. Thus the polynomiality property of Eq.~(\ref{En}), namely that the
$X^{n-1}$ moment of $E_{T}^{\pi }\left( X,\xi
,t\right) $ is a polynomial in $\xi $ of the order not higher than $n$, is
immediately evident within our approach.

\subsection{Transversity pion form factors in momentum- and impact-parameter spaces \label{sec:tff}}

>From representation (\ref{MTn}), by using the definition of the tGFFs (\ref%
{PionTme}), one gets\footnote{%
In the following we will explore the strict chiral limit of $m_{\pi }=0.$}
\begin{align}
& B_{Tni}^{u}\left( t\right) =\frac{N_{c}}{4\pi ^{2}f_{\pi }^{2}}\frac{%
\left( n-1\right) !}{i!\left( n-1-i\right) !}\int \frac{d\left( \alpha \beta
\gamma \right) }{\Delta ^{n+2}}e^{-\frac{\alpha \gamma }{\Delta }t}
\label{BTn} \\
& \left[ 2\alpha G_{m,0,0}\left( \alpha ,\beta ,\gamma \right) +\beta
G_{0,m,0}\left( \alpha ,\beta ,\gamma \right) \right] \beta ^{n-1-i}\left(
\frac{\gamma -\alpha }{2}\right) ^{i}\!\!,  \notag
\end{align}%
where $i=0,2,...\leq n-1$, and the symmetry properties under the interchange of $\alpha $ and $\gamma $
has been used. The transverse (impact parameter) space representation is obtained, by definition,
after a 2D Fourier-Bessel transformation,
\begin{equation}
F\left( b_{\perp }^{2}\right) =\int \frac{d^{2}q_{\perp }}{\left( 2\pi
\right) ^{2}}e^{-i(b_{\perp }q_{\perp })}F\left( t=-q_{\perp }^{2}\right).
\label{2dFourier}
\end{equation}%
We then get for even $i$ the expression
\begin{align}
& B_{Tni}^{u}\left( b_{\perp }^{2}\right) =\frac{N_{c}}{16\pi ^{3}f_{\pi
}^{2}}\frac{\left( n-1\right) !}{i!\left( n-1-i\right) !}\int \frac{d\left(
\alpha \beta \gamma \right) }{\alpha \gamma \Delta ^{n+1}}e^{-\frac{\Delta }{%
\alpha \gamma }\frac{b_{\perp }^{2}}{4}}  \notag \\
& \left[ 2\alpha G_{m,0,0}\left( \alpha ,\beta ,\gamma \right) +\beta
G_{0,m,0}\left( \alpha ,\beta ,\gamma \right) \right] \beta ^{n-1-i}\left(
\frac{\gamma -\alpha }{2}\right) ^{i}\!\!. \label{BTnTr}
\end{align}

\subsection{Pion transversity Generalized Parton Distribution \label{sec:tgpd}}

Through the use of the definition of the tGPD in Eq.~(\ref{En}) we arrive at the formula
\begin{align}
& E_{T}^{\pi }\left( X,\xi ,t\right) =\frac{N_{c}}{4\pi ^{2}f_{\pi
}^{2}}\int \frac{d\left( \alpha \beta \gamma \right) }{\Delta ^{3}}e^{-\frac{%
\alpha \gamma }{\Delta }t} \times  \label{ETn} \\
& \left[ \alpha G_{m,0,0}\left( \alpha ,\beta ,\gamma \right) +\beta
G_{0,m,0}\left( \alpha ,\beta ,\gamma \right) +\gamma G_{0,0,m}\left( \alpha
,\beta ,\gamma \right) \right]   \notag \\
& \times \delta \left( X-\frac{\beta +\left( \gamma -\alpha \right) \xi }{%
\Delta }\right) ,  \notag \\
&  -1<X=\frac{\beta +\left( \gamma -\alpha \right) \xi }{\Delta }<1.
\notag
\end{align}%
Let us integrate over the $\beta $ parameter, corresponding to the quark
spectator. From the $\delta $ function we resolve $\beta $ as%
\begin{equation}
\beta =\frac{ \left( X+\xi \right) \alpha +\left( X-\xi \right) \gamma }{1-X}  \label{beta}
\end{equation}%
and apply the positivity conditions for $\alpha$, $\beta$, and $\gamma$.
At fixed $\xi \in  \left[ 0,1\right] $
and $X \in \left[ -1,1\right] $ one has 3 distinct regions:
\begin{align*}
& \mathrm{I}.\quad \xi <X<1,\quad \mathrm{where}\quad X%
+\xi >0,X-\xi >0, \\
& \mathrm{II}.\quad -\xi <X<\xi ,\quad \mathrm{where}\quad \mathrm{X%
}+\xi >0,X-\xi <0, \\
& \mathrm{III}.\quad -1<X<-\xi ,\quad \mathrm{where}\quad X%
+\xi <0,X-\xi <0.
\end{align*}%
In region I $\beta $ is positive without any limitations. In region
III all coefficients in Eq.~(\ref{beta}) are negative, hence the support of the integrand
has zero measure and the integral in Eq.~(\ref{ETn}) equals zero. In
the central region II the coefficient of $\alpha $ in Eq.~(\ref{beta}) is
positive and the coefficient of $\gamma $ is negative, thus one has
the limitation $\alpha >\gamma \frac{\xi -X}{\xi +X}$. Finally, the
total result may be combined as%
\begin{align}
& E_{T}^{\pi }\left( X,\xi ,t\right) =\Theta \left( X+\xi
\right) \frac{N_{c}}{4\pi ^{2}f_{\pi }^{2}}\int_{0}^{\infty } \!\!\!\!\! d\gamma
\int_{{\rm max}\left\{ 0,\gamma \frac{\xi -X}{\xi +X}\right\}
}^{\infty } \!\!\!\!\!\!\!\!\!\!\!\!\!\!\!\!\!\!\! d\alpha \, e^{-\frac{\alpha \gamma }{\Delta }t} \times \notag \\
& \frac{\alpha G_{m,0,0}\left( \alpha ,\beta ,\gamma \right) +\beta
G_{0,m,0}\left( \alpha ,\beta ,\gamma \right) +\gamma G_{0,0,m}\left( \alpha
,\beta ,\gamma \right) }{\Delta ^{2}\left( 1-X\right) },  \label{ETnM}
\end{align}%
where  $\Theta\left( x\right) $ is the step function, $\beta $ is given by Eq.~(\ref{beta}), and
$\Delta =\left[ \alpha +\gamma +\xi \left( \alpha -\gamma \right) \right]/({1-X})$.

The isovector and isosinglet tGPDs of the pion are obtained as the symmetric and antisymmetric
combinations,%
\begin{align}
E_{T}^{\pi ,I=1}\left( X,\xi ,Q^{2}\right) & \equiv E_{T}^{\pi ,S}\left( X,\xi ,Q^{2}\right) \notag \\ &= E_{T}^{\pi }\left(
X,\xi ,Q^{2}\right) +E_{T}^{\pi }\left( -X,\xi,Q^{2}\right) ,   \notag \\
E_{T}^{\pi ,I=0}\left( X,\xi ,Q^{2}\right) & \equiv E_{T}^{\pi ,A}\left( X,\xi ,Q^{2}\right) \notag \\ &= E_{T}^{\pi }\left(
X,\xi ,Q^{2}\right) -E_{T}^{\pi }\left( -X,\xi,Q^{2}\right) .  \label{ETnI01}
\end{align}
The support of $E_{T}^{\pi ,I=0,1}$ is $-1 \le X \le 1$. The significance of the isospin combinations comes from the fact that
they evolve autonomously with the renormalization scale, see Sec.~\ref{sec:evol}.

\subsection{Special kinematics: $\xi =0$ and $\xi=X$ cases \label{sec:special}}

Some special kinematics is evident. For the case $\xi =0$ (tPDF) we have
\begin{align}
& E_{T}^{\pi }\left( X,\xi =0,t\right) =\Theta \left( X%
\right) \frac{N_{c}}{4\pi ^{2}f_{\pi }^{2}}\int_{0}^{\infty }d\left( \alpha
\gamma \right) e^{-\frac{\alpha \gamma }{\Delta }t} \times \notag  \\
& \frac{2\alpha G_{m,0,0}\left( \alpha ,\beta ,\gamma \right) +\beta
G_{0,m,0}\left( \alpha ,\beta ,\gamma \right) }{\Delta ^{2}\left( 1-\mathrm{X%
}\right) },  \label{ETnMF}
\end{align}%
where $\beta =\left( \alpha +\gamma \right) \frac{X}{1-X}$ and
$\Delta =\left( \alpha +\gamma \right) \frac{1}{1-X}$. Note that in general
the first term in the numerator dominates in the small $X$ region, while the second one
is more important in the region of large $X$.

For the border case,  $\xi=X$, we find
\begin{align}
& E_{T}^{\pi }\left( X,\xi=X,t\right) =\Theta\left(
X\right) \frac{N_{c}}{4\pi ^{2}f_{\pi }^{2}}\int_{0}^{\infty
}d\left( \alpha \gamma \right) e^{-\frac{\alpha \gamma }{\Delta }t}
\times \notag \\
& \frac{\alpha G_{m,0,0}\left( \alpha ,\beta ,\gamma \right) +\beta
G_{0,m,0}\left( \alpha ,\beta ,\gamma \right) +\gamma G_{0,0,m}\left( \alpha
,\beta ,\gamma \right) }{\Delta ^{2}\left( 1-X\right) }, \label{ETnMD}
\end{align}%
with $\beta =2\alpha \frac{X}{1-X}$ and $\Delta =\left[ \alpha
+\gamma +X\left( \alpha -\gamma \right) \right] \frac{1}{1-X}$.

\subsection{Double Distribution \label{sec:DD}}

Some symmetry properties of the GPDs are more transparent when they
are constructed from the double distributions (DDs) \cite%
{Mueller:1998fv,Radyushkin:1996nd,Radyushkin:2011dh}. Actually, the relativistic invariance
exhibited by the polynomiality conditions is manifestly built-in in this approach (see,
e.g., Ref.~\cite{BAG}). To pass to double distributions, we first make the substitution (see, e.g., \cite{Radyushkin:2011dh}) $%
\alpha =x_{1}L,\beta =x_{2}L,\gamma =x_{3}L$ in Eq.~(\ref{ETn}) and obtain%
\begin{align}
& E_{T}^{\pi }\left( X,\xi ,t\right) =\frac{N_{c}}{4\pi ^{2}f_{\pi
}^{2}}\int_{0}^{\infty }dL\int_{0}^{1}dx_{1}dx_{2}dx_{3}e^{-x_{1}x_{3}t} \times \notag \\
& \delta \left( 1-x_{1}-x_{2}-x_{3}\right) \delta \left( x-x_{2}-\left(
x_{3}-x_{1}\right) \xi \right) \times \notag \\
& \left[ x_{1}G_{m,0,0}\left( x_{1}L,x_{2}L,x_{3}L\right)
+x_{2}G_{0,m,0}\left( x_{1}L,x_{2}L,x_{3}L\right) \right. \notag \\
& \left. +x_{3}G_{0,0,m}\left( x_{1}L,x_{2}L,x_{3}L\right) \right] .
\end{align}%
To recover the DD representation we further make the replacement $%
x_{2}=b,x_{3}-x_{1}=a$ and arrive at%
\begin{equation}
E_{T}^{\pi }\left( X,\xi ,t\right)
=\int_{0}^{1}db\int_{-1+b}^{1-b}da\delta \left( X-b-a\xi \right)
f_{T}^{\pi }\left( a,b,t\right) ,  \label{ETnDD}
\end{equation}%
with the DD identified as%
\begin{align}
& f_{T}^{\pi }\left( a,b,t\right) =\frac{N_{c}}{4\pi ^{2}f_{\pi }^{2}}%
\int_{0}^{\infty }dL\,e^{-x_{1}x_{3}t} \times  \label{DD} \\
& \left[ x_{1}G_{m,0,0}\left( x_{1}L,bL,x_{3}L\right) +bG_{0,m,0}\left(
x_{1}L,bL,x_{3}L\right) \right.   \notag \\
& \left. +x_{3}G_{0,0,m}\left( x_{1}L,bL,x_{3}L\right) \right] .  \notag
\end{align}%
Here $x_{1}=\frac{1}{2}\left( 1-b-a\right)$ and $x_{3}=\frac{1}{2}\left(
1-b+a\right) $. In the above expressions the parameter $b$ is
non-negative. The $b\leq 0$ part of the DD comes from the crossed diagram.

Sometimes it is also convenient to separate the so-called D-term, defined as%
\begin{equation}
D\left( b,t\right) =\int_{-1+b}^{1-b}daf_{T}^{\pi}\left( a,b,t\right) .
\label{Dterm}
\end{equation}

\subsection{The $b_{\perp}$ space and the transverse pion size \label{sec:bperp}}

Let us now consider tGPD in the transverse coordinate space, $b_{\perp}$. By using the
2D Fourier-Bessel transform of Eq.~(\ref{2dFourier}) one easily gets%
\begin{align}
& E_{T}^{\pi }\left( X,\xi ,b_{\perp }^{2}\right) =\Theta \left(
X+\xi \right) \frac{N_{c}}{16\pi ^{3}f_{\pi }^{2}} 
\int_{0}^{\infty } \!\!\!\!\! d\gamma \int_{{\rm max}\left\{ 0,\gamma \frac{\xi -X}{%
\xi +X}\right\} }^{\infty } \hspace{-1.75cm} d\alpha \, e^{-\frac{\Delta }{\alpha \gamma
}\frac{b_{\perp }^{2}}{4}} \times \notag \\
& \frac{\alpha G_{m,0,0}\left( \alpha ,\beta ,\gamma \right) +\beta
G_{0,m,0}\left( \alpha ,\beta ,\gamma \right) +\gamma G_{0,0,m}\left( \alpha
,\beta ,\gamma \right) }{\Delta \alpha \gamma \left( 1-X\right) },
\end{align}%
where the value of the parameter $\beta $ is given by Eq.~(\ref{beta}) and \mbox{$\Delta =\left[ \alpha +\gamma +\xi
\left( \alpha -\gamma \right) \right] \frac{1}{1-X}$}.

In the zero longitudinal momentum transfer limit, $\xi \rightarrow 0$,
one obtains the so-called 3D transverse parton distribution%
\begin{equation}
f_{T}^{\pi }\left( X,b_{\perp }\right) =E_{T}^{\pi }\left( \mathrm{X%
},\xi \rightarrow 0,b_{\perp }^{2}\right) .
\end{equation}%
Following \cite{Perevalova:2011qi} one can also introduce the normalized
quark probability density in the transverse plane,%
\begin{equation}
\rho _{T}^{\pi }\left( X,b_{\perp }\right) =\frac{f_{T}^{\pi
}\left( X,b_{\perp }\right) }{f_{T}^{\pi }\left( X\right) }%
,  \label{QPD}
\end{equation}%
where
\begin{eqnarray}
f_{T}^\pi \left( X\right) \equiv E_{T}^{\pi }\left( X,\xi =0,t=0\right),
\label{fTpi}\end{eqnarray}
as defined in (\ref{ETnMF}). The partons with the
longitudinal momentum fraction $X$ occupy within the hadron a disc of
the average transverse radius squared given by
\begin{equation}
b_{\perp }^{2}\left( X\right) =\int d^{2}b_{\perp }b_{\perp
}^{2}f_{T}^{\pi }\left( X,b_{\perp }\right) .  \label{b2x}
\end{equation}%

In chiral quark models the triangle diagram yields
\begin{align}
& b_{\perp }^{2}\left( X\right) =\frac{N_{c}}{\pi ^{2}f_{\pi }^{2}}%
\left( 1-X\right) ^{2}\int d\left( \alpha \gamma \right) \frac{%
\alpha \gamma }{\left( \alpha +\gamma \right) ^{3}} \times \label{b2xModel} \\
& \left[ 2\alpha G_{m,0,0}\left( \alpha ,\beta ,\gamma \right) +\beta
G_{0,m,0}\left( \alpha ,\beta ,\gamma \right)  \right] ,  \notag
\end{align}%
where $\beta =\left( \alpha +\gamma \right) \frac{X}{1-X}$%
. The C-odd transverse size of the hadron, determined by the slope
of the tGFF at low momentum transfer, can be obtained by integrating $b_{\perp
}^{2}(X)$ over the momentum fraction,
\begin{equation}
b_{\perp }^{2}=2\int_{0}^{1}dX\,b_{\perp }^{2}\left( X \right) .  \label{b2}
\end{equation}

According to Gribov~\cite{Gribov:1973jg}, one can interpret the
normalized quark density (\ref{QPD}) as an evolution of the
probability density for a stochastic motion of a particle in the
transverse plane. The role of the evolution time is played by the
rapidity variable, $\eta =\ln(1/X)$. For the stochastic process one can
introduce the mean squared distance of the particle as follows
\cite{Perevalova:2011qi}:%
\begin{equation}
d_{\perp }^{2}\left( X\right) =\int d^{2}b_{\perp }b_{\perp
}^{2}\rho \left( X,b_{\perp }\right) =\frac{b_{\perp }^{2}\left(
X\right) }{f\left( X\right) }.  \label{D2T}
\end{equation}%
By using a model with short-range interactions, Gribov predicted that
\cite{Gribov:1973jg}%
\begin{equation}
d_{\perp }^{2}\left( \eta \right) =D\eta ,  \label{D2Tgribov}
\end{equation}%
where $D$ is a constant,
while in \cite{Perevalova:2011qi} the result is%
\begin{equation}
d_{\perp }^{2}\left( \eta \right) \sim \frac{1}{\left( 4\pi f_{\pi }\right)
^{2}}e^{\left( 1-\omega \right) \eta }.  \label{D2Tpolyakov}
\end{equation}%
Here $\omega \approx 0.5$ is the slope of the forward quark
distribution at small $X$, i.e., $q(X)\sim 1/X^{\omega }$.
Note that Eq.~(\ref{D2Tpolyakov}) is
${\cal O} (N_c^{-1})$, since $f_\pi = {\cal O}(\sqrt{N_c}) $. Actually,
the ``chiral inflation'' discussed in Ref.~\cite{Perevalova:2011qi} is
a pion-loop effect, which is $1/N_c$-suppressed, but at the same time it is chirally
enhanced as $\log ( m_\pi^2)$ for $m_\pi \to 0$, compared to the
leading one-quark-loop contribution. In the real world with $N_c=3$
and $m_\pi=140{\rm ~MeV}$ the relative chiral contributions to the rms radius
of the pion are about 20\%~\cite{Gasser:1983yg}~\footnote{Actually,
from the relation for the rms radius of the pion found in ChPT~\cite{Gasser:1983yg},
$\langle r^2 \rangle = (\bar l_5 -1)/(16\pi^2 f_\pi^2)$, one has the
{\it total} low energy constant $\bar l_5=13.9\pm 1.3$, most of which
is saturated by the $\rho$-meson exchange, $\bar l_5^\rho \simeq 17$, at the
leading order in $N_c$. Thus, the subleading ($1/N_c $-suppressed)
contribution is estimated to be $\Delta \bar l_5 \equiv \bar l_5 -
\bar l_5^\rho \sim \log(m_\pi^2/m_\rho^2)\sim - 3$.}. Of course, the
additional inclusion of pion-loops in our model would automatically
reproduce this universal inflating phenomenon.

\section{Model results \label{sec:specific}}

Having derived the general formulas for tGPDs in chiral quark models from the triangle diagram of Fig. \ref{fig:tri}, we now pass
to presenting explicit numerical calculations. We start with the nonlocal models.
In the present work we consider two variants of the quark-pion
vertex of Eq.~(\ref{QPiVert}),
\begin{align}
& F_{\mathrm{I}}\left( k_{+}^{2},k_{-}^{2}\right) =\sqrt{m\left(
k_{+}^{2}\right) m\left( k_{-}^{2}\right) },  \label{QPiVertI} \\
& F_{\mathrm{HTV}}\left( k_{+}^{2},k_{-}^{2}\right) =\frac{1}{2}\left[
m\left( k_{+}^{2}\right) +m\left( k_{-}^{2}\right) \right] ,
\label{QPiVertT}
\end{align}
where $m\left( k^{2}\right) $ is the momentum-dependent dynamical quark
mass.  The form (\ref{QPiVertI}) is motivated by the instanton picture
of the QCD vacuum \cite{Diakonov:1985eg} and is labeled ``instanton'', while the form (\ref{QPiVertT}), the
Holdom-Terning-Verbeek (HTV) vertex, comes from the nonlocal chiral
quark model of Ref.~\cite{Holdom:1990iq}. Some relevant differences between
both prescriptions regarding the proper implementation of chiral
symmetry are discussed in Ref.~\cite{Broniowski:1999dm}.

We consider the dynamical quark mass of the form%
\begin{equation}
m\left( k^{2}\right) =M_{q}f^{2}\left( k^{2}\right) ,
\end{equation}%
and for simplicity take the profile function $f(k^{2})$ as a Gaussian,
\begin{equation}
f(k^{2})=e^{-\Lambda k^{2}}  \label{GaussF}
\end{equation}%
(note that $\Lambda$ has the interpretation of the squared inverse momentum cut-off).
The model contains two parameters: the dynamical quark mass at zero momentum, $M_{q}$, and
the nonlocality scale, $\Lambda $. For our numerical estimates we take one
parameter fixed at a physically reasonable value, $M_{q} \simeq 240$~MeV, and then fix $\Lambda $
via the pion decay constant evaluated in the chiral limit, $f_{\pi }=84$~MeV \cite{Gasser:1983yg}.
The expression for $f_{\pi }$ in the instanton model is given by
the Diakonov-Petrov formula \cite{Diakonov:1985eg},
\begin{eqnarray}
&&\hspace{-4mm} f_{\pi }^{\mathrm{I}}=\left[ \frac{N_{c}}{4\pi ^{2}}\! \int_{0}^{\infty
} \!\!\!\!\! du \, u\frac{m\left( u\right) }{D^{2}\left( u\right) } \left( m\left( u\right)\! - \!um^{\prime }\left( u\right)
\!+ \! u^{2}m^{\prime 2}( u) \right) \right] ^{1/2} \!\!\!\!, \nonumber \\
&&\label{FpiDP}
\end{eqnarray}%
while in the HTV model one has the Pagels-Stokar formula \cite%
{Pagels:1979hd,Holdom:1990iq}
\begin{equation}
f_{\pi }^{\mathrm{HTV}}=\left[ \frac{N_{c}}{4\pi ^{2}}\int_{0}^{\infty
}du\, u\frac{m\left( u\right) }{D^{2}\left( u\right) }\left( m\left(
u\right) -\frac{1}{2}um^{\prime }\left( u\right) \right) \right] ^{1/2}.
\label{FpiPS}
\end{equation}%
The described parameter-fitting procedure yields
\begin{equation}
\Lambda _{\mathrm{I}}=0.7~\mathrm{GeV}^{-2},\quad \Lambda _{\mathrm{HTV}%
}=0.375~\mathrm{GeV}^{-2}.
\end{equation}

For the instanton model, the integrand in Eq.~(\ref{MTn}) and the subsequent
formulas can be expressed as follows:
\begin{eqnarray}
&&\alpha G_{m,0,0}\left( \alpha ,\beta ,\gamma \right) +\beta
G_{0,m,0}\left( \alpha ,\beta ,\gamma \right) +\gamma G_{0,0,m}\left( \alpha
,\beta ,\gamma \right)   \notag \\
&&\overset{\mathrm{I}}{\rightarrow }\alpha d_{\alpha }^{3/2}d_{\beta
}^{1} d_{\gamma }^{1/2}+\beta d_{\alpha }^{1/2} d_{\beta }^{2} d_{\gamma
}^{1/2}+\gamma d_{\alpha }^{1/2} d_{\beta }^{1} d_{\gamma }^{3/2},  \label{II}
\end{eqnarray}%
while for the HTV model one has%
\begin{align}
& \alpha G_{m,0,0}\left( \alpha ,\beta ,\gamma \right) +\beta
G_{0,m,0}\left( \alpha ,\beta ,\gamma \right) +\gamma G_{0,0,m}\left( \alpha
,\beta ,\gamma \right) \notag \\
& \overset{\mathrm{HTV}}{\rightarrow }\frac{1}{4}\left\{ \alpha \left(
d_{\alpha }^{2}d_{\beta }^{1}d_{\gamma }^0+d_{\alpha }^{2}d_{\beta
}^0 d_{\gamma }^{1}+d_{\alpha }^{1}d_{\beta }^{2}d_{\gamma }^0+d_{\alpha
}^{1}d_{\beta }^{1}d_{\gamma }^{1}\right) \right.   \notag \\
& +\left( \alpha \longleftrightarrow \gamma \right)   \notag \\
& +\beta \left( d_{\alpha }^{1}d_{\beta }^{2}d_{\gamma }^0 +d_{\alpha
}^0 d_{\beta }^{2}d_{\gamma }^{1}+d_{\alpha }^0 d_{\beta }^{3}d_{\gamma
}^0+d_{\alpha }^{1}d_{\beta }^{1}d_{\gamma }^{1}\right) .   \label{IT}
\end{align}%
Here we have introduced the short-hand notation
\begin{align}
\frac{m^{2n}\left(k^{2}\right) }{D\left( k^{2}\right) }\sim d_{\alpha }^{n}.   \label{AlphaExact}
\end{align}

For the assumed Gaussian form factor (\ref{GaussF}) the $d_{\alpha }^{n}$ function at large $%
\alpha \gg \Lambda $ has the following behavior%
\begin{equation}
\frac{1}{R\left( \lambda \right) }M_{q}^{n}e^{-\lambda \left( \alpha
-2n\Lambda \right) }\Theta \left( \alpha -2n\Lambda \right) ,
\label{AlphaApprox}
\end{equation}%
with
\begin{equation}
R\left( \lambda \right) =1-4\Lambda m^{2}\left( \lambda \right) ,
\end{equation}
where $\lambda $ is the root of the equation%
\begin{equation}
\lambda +m^{2}\left( \lambda \right) =0.
\end{equation}%
The functions (\ref{AlphaApprox}) can also be used as approximants for
the analytic calculations of the quark distributions in the pion. In the momentum
representation this simplification means that in the denominators of the
integrands we neglect the momentum dependence of the dynamical quark mass, as would be
the case of the local quark models.

\subsection{The numerical results for nonlocal models \label{sec:NonLocRes}}

\begin{figure}[tb]
\includegraphics[width=.48\textwidth]{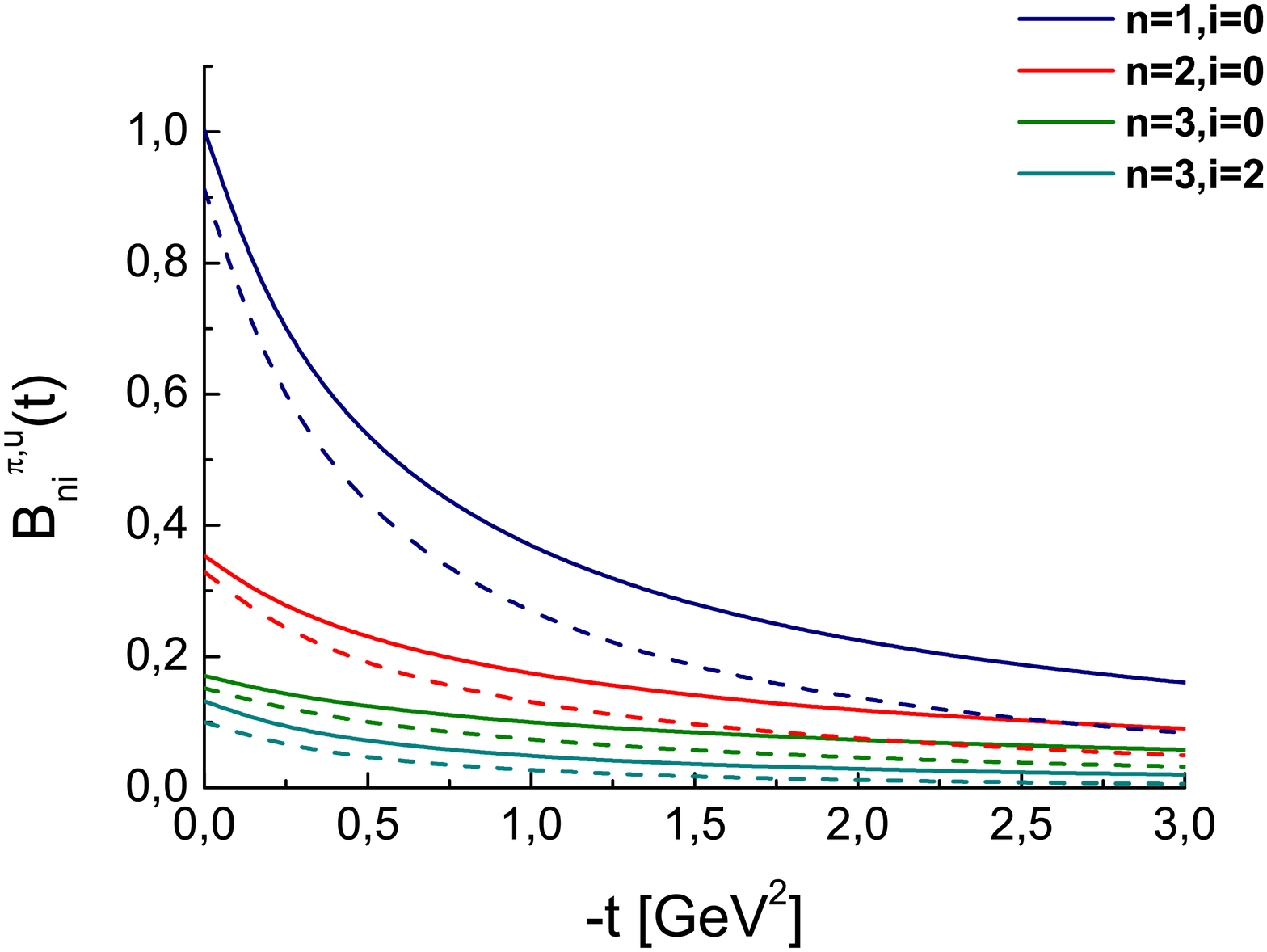}
\caption{(Color online) The tGFFs $B_{ni}^{\pi, u}(t)$ in the HTV model (solid line) and in the
instanton model (dashed line) for several lowest values of $n$ and $i$.
The sequence in the legend corresponds to the sequence of the curves, from top to bottom.
\label{BTniNonLoc}}
\end{figure}

In this subsection we present the results for the nonlocal models.
These results are obtained from the formulas presented above with the help of
numerical integration.

We start by exploring the $t$-dependence. In Fig.~\ref{BTniNonLoc} we present the
pion $u$-quark tGFFs in the HTV model and in the instanton model. First of all, the increase of the indices
$n$ or $i$ causes a decrease of the form factor normalization. We also note a faster fall-off
with $t$ of the tGFFs for the case of the instanton model compared to the HTV case. We note that
the tGFFs undergo the QCD evolution, which will be discussed in detail in Sec.~\ref{sec:evol}.
The $B_{n0}^{\pi, u}$ form factors, however, evolve multiplicatively, hence we can read off
their $t$-dependence from Fig.~\ref{BTniNonLoc}.

At large $t$ the $B_{10}^{\pi, u}$ form factor in the HTV model has the asymptotic behavior $\sim \ln t/t$.
This follows from the asymptotic formula
\begin{eqnarray}
&&B_{T10}^{u}\left( t \gg \Lambda^{-1} \right) \overset{\mathrm{HTV}}{=}\frac{1}{t}%
\frac{N_{c}}{16\pi ^{2}f_{\pi }^{2}}\left[ \int_{0}^{\infty }du\frac{%
m^{3}\left( u\right) }{D\left( u\right) }\ln \left( \frac{t}{u}\right)
\right. \notag  \\
&&\left. +2\int_{0}^{\infty }du\frac{m^{2}\left( u\right) }{D\left( u\right)
}\int_{0}^{\infty }dv\frac{m\left( u+v\right) }{D\left( u+v\right) } \times \right . \notag \\ && \hspace{2.1cm} \left . \left( 1-%
\frac{m\left( u\right) m\left( u+v\right) }{u+v}\right) \right] .
\end{eqnarray}%
For the instanton model the fall-off is exponential, since
\begin{eqnarray}
&&B_{T10}^{u}\left( t \gg \Lambda^{-1} \right) \overset{\mathrm{I}}{=}\frac{N_{c}}{%
4\pi ^{2}f_{\pi }^{2}}\frac{\sqrt{\pi }M_{q}^{3}}{R\left( \lambda \right) }
\\
&&\frac{1}{t}\frac{1}{\sqrt{\Lambda \sqrt{\lambda t}}}\left( 1-2\sqrt{\frac{%
\lambda }{t}}\right) e^{-\Lambda \left( \sqrt{\lambda t}-6\lambda \right)
}E_{1}\left( \Lambda \sqrt{\lambda t}\right) .  \notag
\end{eqnarray}%

\begin{figure}[tb]
\includegraphics[width=.48\textwidth]{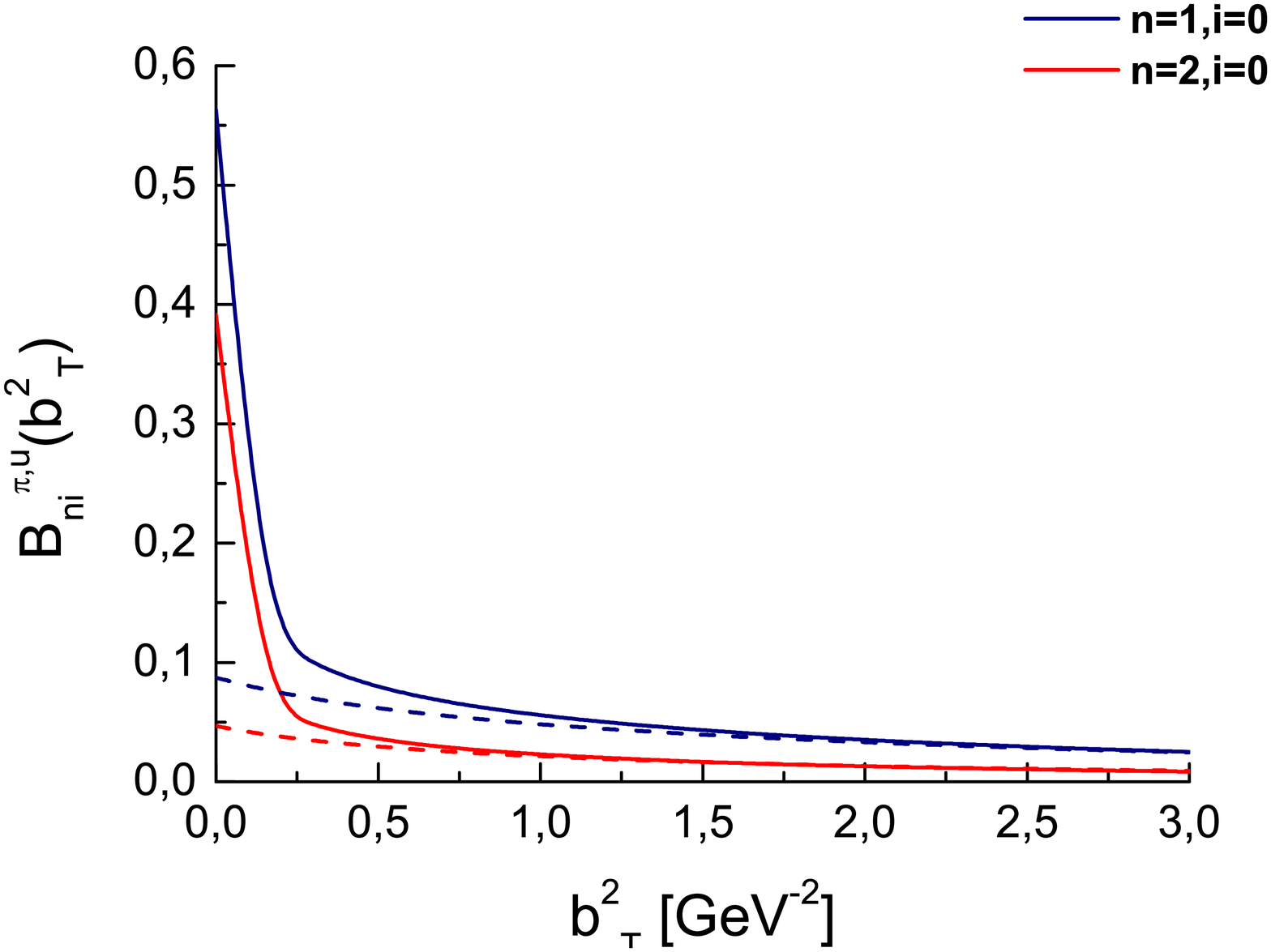}
\caption{(Color online) The tGFFs $B_{ni}^{\pi, u}(b_T^2)$ in the impact parameter space in the HTV
model (solid line) and in the instanton model (dashed line).
The sequence in the legend corresponds to the sequence of the curves, from top to bottom.
\label{BTni(b2)}}
\end{figure}

In Fig.~\ref{BTni(b2)} we display the tGFFs in the impact-parameter space.
The information is the same as in Fig.~\ref{BTniNonLoc}, as the
two figures are simply linked with a Fourier-Bessel transform. Nevertheless, the different
large-$t$ behavior of the instanton and HTV models is very vividly seen in the small-$b_T$ behavior in
Fig.~\ref{BTni(b2)}.

\begin{figure}[tb]
\includegraphics[width=.48\textwidth]{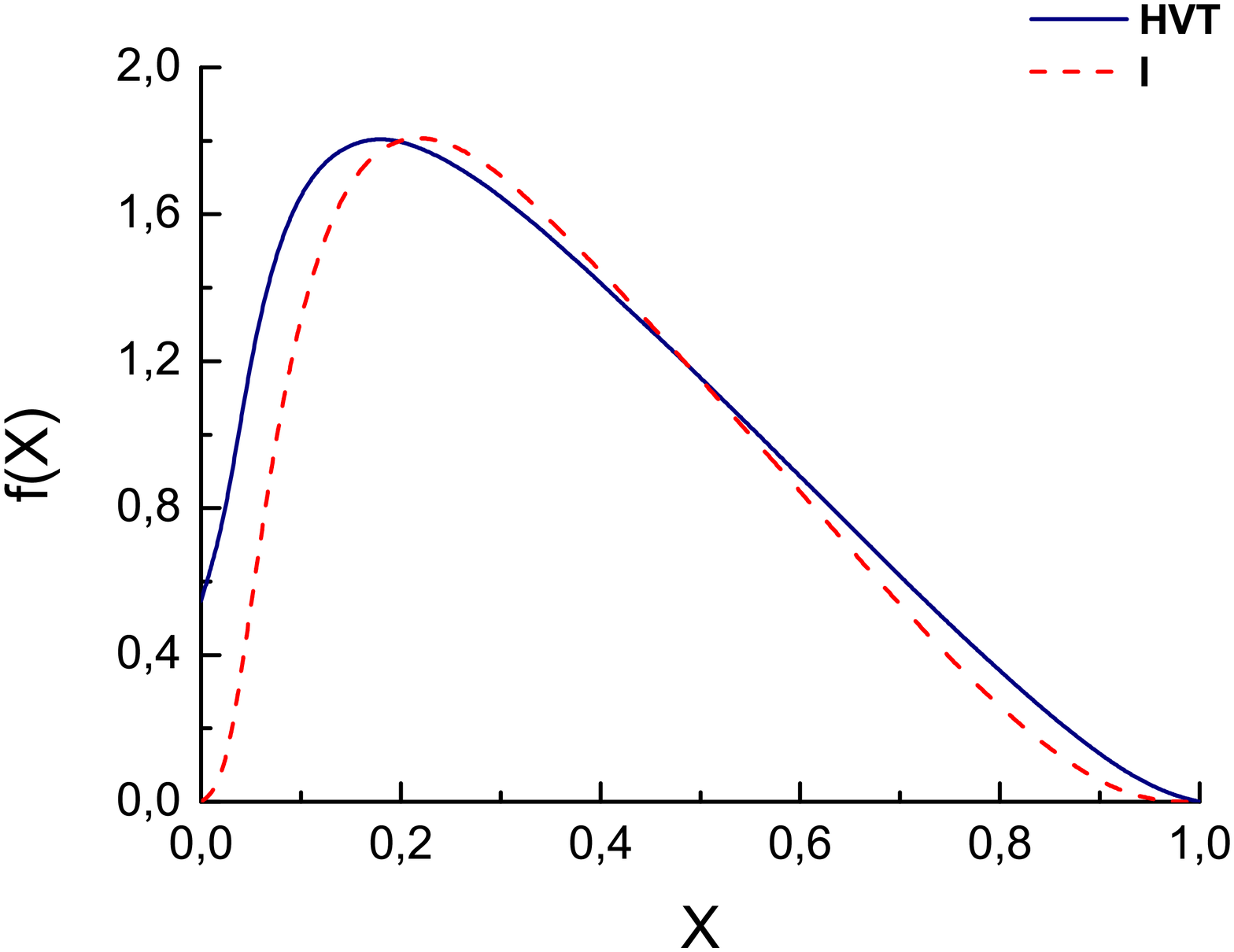}
\caption{(Color online) The tPDF in the HTV model (solid line) and in the
instanton model (dashed line).
\label{f(x)}}
\end{figure}

Next, we explore the $X$ dependence in the simplest case of $t=0$ and $\xi=0$ (tPDF).
In Fig.~\ref{f(x)} we present the results of calculations of the tPDF in
the nonlocal models (\ref{fTpi}).
We notice a more-less triangular shape for both models, with a depletion near $X=0$.

The end-point behavior of these functions can be
inferred from Eq.~(\ref{ETnMF}) by using the approximants (\ref{AlphaApprox}).
The $X\rightarrow 1$ behavior is governed by the properties of the active
dynamical quark, while the $X\rightarrow 0$ behavior is related to the
spectator quark. For the instanton model the endpoint behavior is
exponentially suppressed, namely
\begin{eqnarray}
f_{T}^{I}\left( X\rightarrow 1\right)  &\sim &\left( 1-X\right) ^{2}\exp
\left[ -\frac{2\lambda \Lambda }{1-X}\right] ,  \notag  \\
f_{T}^{I}\left( X\rightarrow 0\right)  &\sim &\exp \left[ -\frac{2\lambda
\Lambda }{X}\right] ,  \label{fI}
\end{eqnarray}%
while for the HVT model one has a power-like behavior%
\begin{eqnarray}
f_{T}^{HVT}\left( X\rightarrow 1\right) &\sim &\left( 1-X\right) , \notag \\
f_{T}^{HVT}\left( X\rightarrow 0\right) &\sim &\mathrm{const.}  \label{fT}
\end{eqnarray}%
We remark here that the end-point behavior in Eqs.~(\ref{fI},\ref{fT}) is
sensitive to the radiative corrections, hence it evolves with the scale.

\begin{figure}[tb]
\includegraphics[width=.48\textwidth]{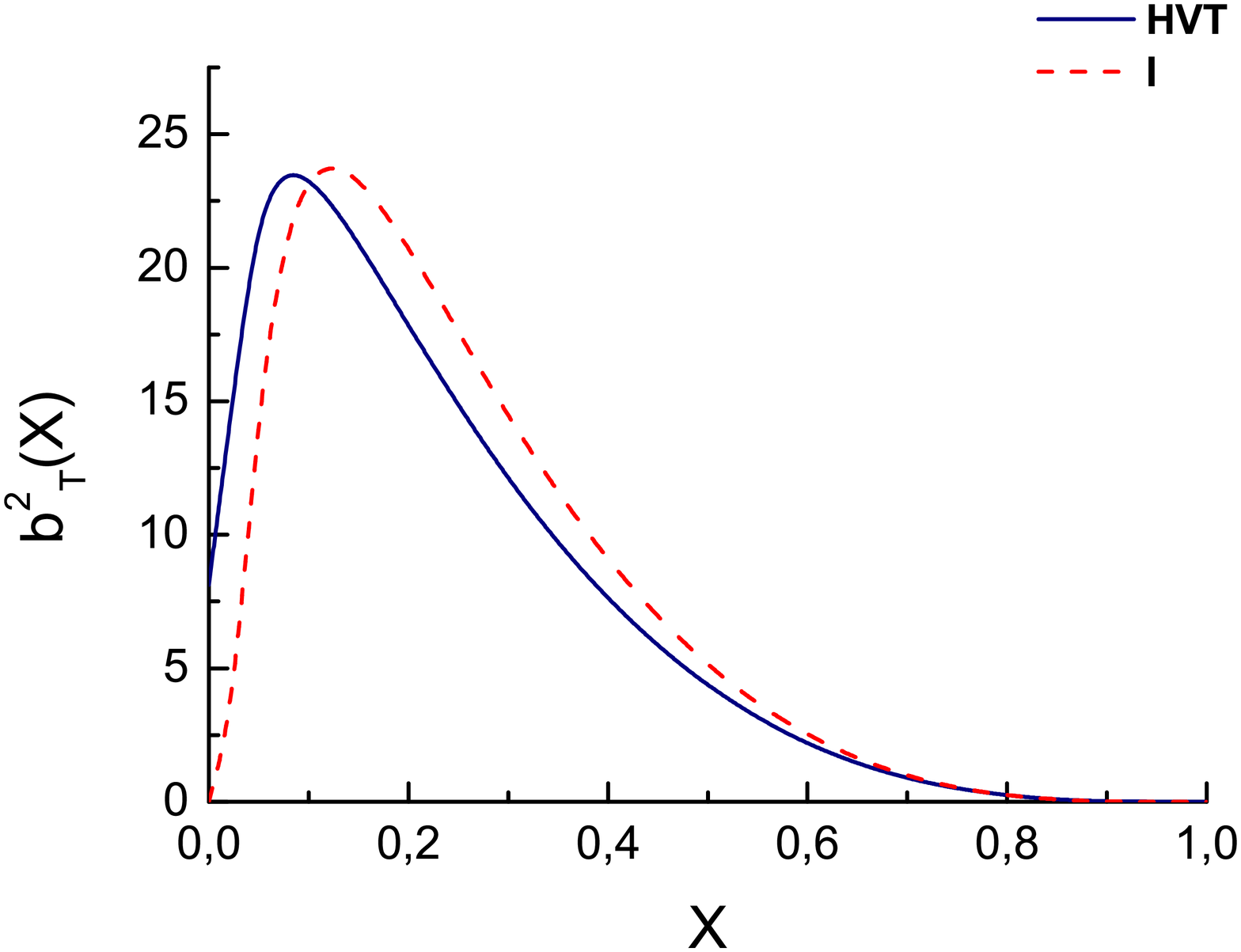}
\caption{(Color online) The distribution function of the transverse size in
the HTV model (solid line) and in the instanton model (dashed
line).
\label{b(x)}}
\end{figure}

A similar behavior is obtained for the transverse size distribution at $t=0$,
shown in Fig. \ref{b(x)}, namely
\begin{eqnarray}
b_{\perp I}^{2}\left( X\rightarrow 1\right)  &\sim &\left( 1-X\right)
^{4}\exp \left[ -\frac{2\lambda \Lambda }{1-X}\right] ,\qquad  \\
b_{\perp I}^{2}\left( X\rightarrow 0\right)  &\sim &\frac{1}{X}\exp \left[ -%
\frac{2\lambda \Lambda }{X}\right] ,  \notag \\
b_{\perp HVT}^{2}\left( X\rightarrow 1\right)  &\sim &\left( 1-X\right)
^{3},\notag \\ b_{\perp HVT}^{2}\left( X\rightarrow 0\right) &\sim &\mathrm{const}.  \notag
\end{eqnarray}

\begin{figure}[tb]
\includegraphics[width=.48\textwidth]{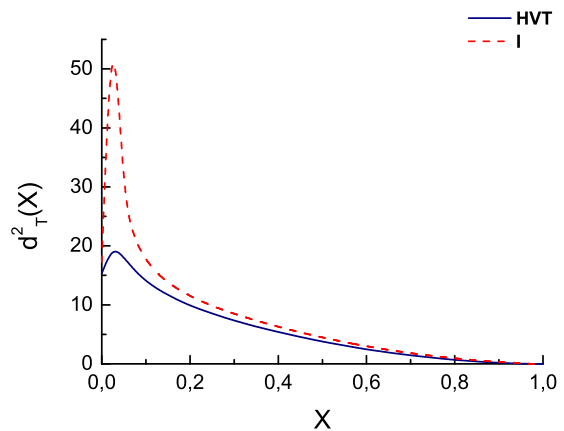}
\caption{(Color online) The distribution function of the mean square
distance in the HTV model (solid line) and in the instanton
model (dashed line), plotted as a function of $X$.
\label{d(x)}}
\end{figure}

\begin{figure}[tb]
\includegraphics[width=.48\textwidth]{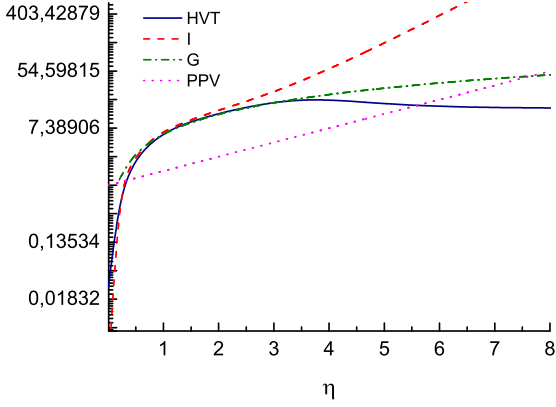}
\caption{(Color online) The distribution function of the mean square
distance as function of rapidity $\protect\eta $ in the HTV model (solid
line), in the instanton model (dashed line), in the Gribov approach \cite{Gribov:1973jg} (G) (dot-dashed line),
and in the PPV model \cite{Perevalova:2011qi} (dotted line).
\label{d(eta)}}
\end{figure}

Next, we present our results for the  distribution function of the mean square
distance.  In Fig.~\ref{d(x)} we show $d_\perp^2$ as a function of $X$, while
in Fig.~\ref{d(eta)} we present the same quantity as a function of the rapidity
variable $\eta$. We also compare our results to the calculations of Refs.~\cite{Gribov:1973jg} (G) and \cite{Perevalova:2011qi} (PPV).
In the region of large $\eta$, corresponding to low $X$, various model predictions are different.

\begin{figure}[tb]
\includegraphics[width=.48\textwidth]{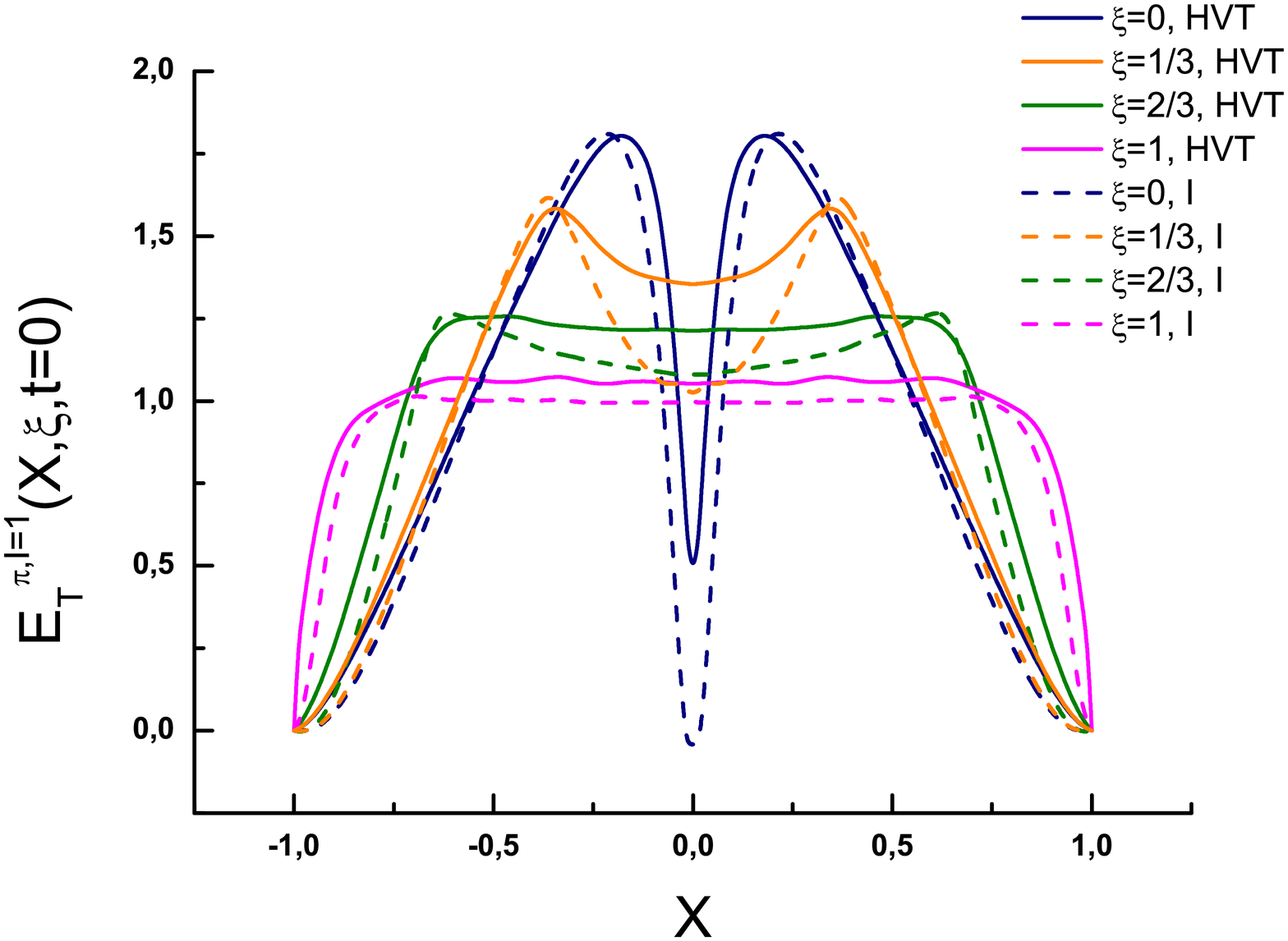}
\caption{(Color online) The pion tGPD for isovector case
in the HTV model (solid lines) and in the instanton model
(dashed lines) for several values of $\xi$.
\label{E(x)I1}}
\end{figure}

\begin{figure}[tb]
\includegraphics[width=.48\textwidth]{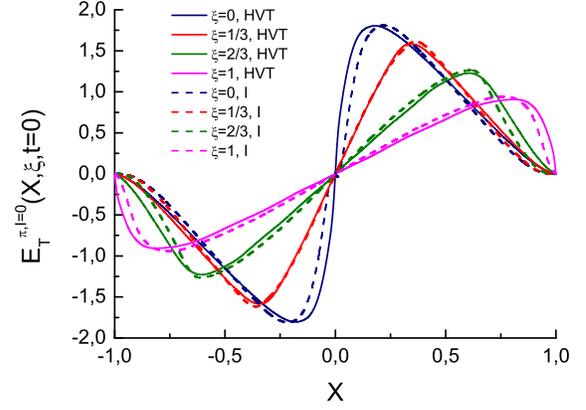}
\caption{(Color online) The pion tGPD for isoscalar case
in the HTV model (solid lines) and in the instanton model
(dashed lines) for several values of $\xi$.
\label{E(x)I0}}
\end{figure}

Finally, we explore the dependence on $\xi$ and $X$ of the pion tGPDs at $t=0$.
The results are given in Figs.~\ref{E(x)I1} and \ref{E(x)I0}.
We note the symmetry properties following from the definition (\ref{ETnI01}).
We can also see that the curves bend near $X=\xi$.

To summarize the study of this subsection we state that the results, apart for
mathematically different end-point behavior, are qualitatively similar in the two
explored variants of the nonlocal chiral quark models.

\subsection{Nambu--Jona-Lasinio model \label{sec:NJL}}

We term the usual Nambu--Jona-Lasinio model with point-like quark-quark
interactions the \emph{local} NJL model. All formulas for the local model
follow from the nonlocal expressions given above, with the constant
quark mass, which formally corresponds to taking the limit $\Lambda \to 0$.
In addition, a regularization prescription, necessary to make the divergent
integrals finite, is implemented, as discussed below.

The one-quark-loop action of the NJL model is
\begin{align}
\Gamma_{\mathrm{NJL}} =-i N_{c} \mathrm{Tr} \log\left( i\setbox0=%
\hbox{$\partial$} \dimen0=\wd0 \setbox1=\hbox{/} \dimen1=\wd1 \ifdim\dimen0>%
\dimen1 \rlap{\hbox to \dimen0{\hfil/\hfil}} \partial \else
\rlap{\hbox to
\dimen1{\hfil$\partial$\hfil}} / \fi - M U^{5} - m \right) \Big|_{\mathrm{reg%
}} ,  \label{eq:eff_ac_NJL}
\end{align}
where $M$ is the constituent quark mass generated via the spontaneous
breaking of the chiral symmetry,
\begin{align}
U^{5}=\exp(i \gamma_{5} \mathchoice{\mbox{\boldmath$\displaystyle \phi$}}
{\mbox{\boldmath$\textstyle \phi$}} {\mbox{\boldmath$\scriptstyle \phi$}}
{\mbox{\boldmath$\scriptscriptstyle \phi$}}\cdot
\mathchoice{\mbox{\boldmath$\displaystyle \tau$}} {\mbox{\boldmath$\textstyle \tau$}} {\mbox{\boldmath$\scriptstyle \tau$}} {\mbox{\boldmath$\scriptscriptstyle
\tau$}}),
\end{align}
with $\mathchoice{\mbox{\boldmath$\displaystyle \phi$}} {\mbox{\boldmath$%
\textstyle \phi$}} {\mbox{\boldmath$\scriptstyle \phi$}} {%
\mbox{\boldmath$\scriptscriptstyle \phi$}}$ denoting the pion field, while $%
m $ is the current quark mass. We apply the NJL with the Pauli-Villars
regularization in the twice-subtracted version of Refs.~\cite%
{RuizArriola:1991gc,Schuren:1991sc, RuizArriola:2002wr}. Variants of chiral
quark models differ in the way of performing the necessary regularization of
the quark loop diagrams, which may to some extent influence the physical
results.

Here we use the prescription where $M^{2}$ in the loop integral is replaced
with the combination $M^{2}+\Lambda^{2}$, where in the present context
$\Lambda$ is the cut-off
parameter, and then the regularized observable is evaluated according to the
formula
\begin{eqnarray}
\mathcal{O}_{\mathrm{reg}} = \mathcal{O}(0) - \mathcal{O}(\Lambda^{2} ) +
\Lambda^{2} d \mathcal{\ O}(\Lambda^{2} ) / d\Lambda^{2}.
\end{eqnarray}
The pre-multiplying factor $g_{\pi}^{2}=M^{2}/f_{\pi}^{2}$ is not
regularized~\cite{RuizArriola:1991gc,Schuren:1991sc, RuizArriola:2002wr}.

In the local model it is relatively simple to go beyond the chiral limit, hence we
do not restrict ourselves to the case $m_\pi=0$.
Since the lattice data used in this work are actually for $m_\pi=600$~MeV, hence not at all
close to the chiral limit of $m=0$, we need to deal with a situation of
moderately large pion masses. The prescription to fix the model parameters
is as follows: the three constants $\Lambda$, $M$, and $m$ are traded for
the constituent quark mass, $M$, the pion decay constant $f_{\pi}$, and $%
m_{\pi}$. We assume that $\Lambda$ depends on $M$ only, and not on $m$.
Constraining $f_{\pi}=93$~MeV (the physical value) and using the given value
of $m_{\pi}$ leaves us with one free parameter only, $M$, which is taken
in the $250-300$~MeV ball park.

We recall that the optimum value of $M$ used in chiral quark
models depends of particular observable used for the fitting procedure. The
application to the $\rho $ meson suggests $M$ above $m_{\rho }/2\sim 400$~MeV, while
the soliton models for the nucleon prefer $M\sim 300-350$~MeV \cite%
{Christov:1995vm}. However, significantly lower values follow from other studies in the
pion sector. The
charge radius of the pion in the NJL model with the Pauli-Villars regulator
favors $M\sim 280$~MeV \cite{RuizArriola:2002wr}, however, the pion-loop
corrections to this observable are important. The analysis of the radii of
the pion charge and transition form factors from quark triangle diagrams
yields $M=\sqrt{2/3}\,\pi f_{\pi }\sim 240$~MeV \cite{Gerasimov:1978cp}.
Another restriction on the value of $M$ follows from the Adler function and the
corresponding vacuum polarization contribution to the gyromagnetic factor $g-2$
of the muon. The loop approach (without and with radiative corrections) \cite%
{Pivovarov:2001mw,Boughezal:2011vw} yields $M=180-200$~MeV, the analytic perturbation model \cite{Milton:2001mq} gives $240$~MeV,
while the nonlocal chiral quark model \cite{Dorokhov:2004ze} suggests $250$~MeV. Our chosen
value of $\sim 250$~MeV falls into this ball park.

In the NJL model the formulas for the lowest two transversity form factors are very simple,
\begin{align}
& \frac{B_{T10}^{\pi ,u}(t)}{m_{\pi }}=\int_{0}^{1}\!\!\!d\alpha
\int_{0}^{1-\alpha }\!\!\!\!\!\!d\beta \,K,
\notag \\
& \frac{B_{T20}^{\pi ,u}(t)}{%
m_{\pi }}=\int_{0}^{1}\!\!\!d\alpha \int_{0}^{1-\alpha }\!\!\!\!\!\!d\beta
\,\alpha K,  \notag \\
& K=\left. \frac{N_{c}g_{\pi }^{2}M}{2\pi ^{2}\left( M^{2}+m_{\pi
}^{2}(\alpha -1)\alpha +t\beta (\alpha +\beta -1)\right) }\right\vert _{%
\mathrm{reg}}.  \label{eq:NJL}
\end{align}%
with $g_{\pi}=M/f_{\pi}$. The variables $\alpha $ and $\beta $ are the Feynman parameters.

The result for the tGPD are particularly simple at $t=0$ and in the chiral limit, namely trapezoidal for
the symmetric ($I=1$) combination,
\begin{equation}
E_{T}^{\pi ,S}(X,\xi ,t=0;\mu _{0})/N=\left\{
\begin{array}{rl}
1, & 0\leq X\leq \xi  \\
\frac{1-X}{1-\xi }, & \xi \leq X\leq 1%
\end{array}%
\right. ,
\end{equation}%
and triangular for the antisymmetric ($I=0$) combination,
\begin{equation}
E_{T}^{\pi ,A}(X,\xi ,t=0;\mu _{0})/N=\left\{
\begin{array}{rl}
X/{\xi }, & 0\leq X\leq \xi  \\
\frac{1-X}{1-\xi }, & \xi \leq X\leq 1%
\end{array}%
\right. .
\end{equation}
Here $N$ denotes a normalization constant following from the model.

Other results of the local NJL model, the corresponding plots, and comparisons to the predictions
of the nonlocal models will be presented in the following parts, together
with the discussion of the QCD evolution.

\section{QCD evolution \label{sec:evol}}

We now come to a very important aspect of our analysis.
Before comparing the results to the lattice data we need to carry out the
QCD evolution, as the tGPD and tGFFs evolve with the scale. The need for the
evolution has been discussed in detail in \cite{Broniowski:2007si}. In
essence, our approach consists of 1)~evaluation of the appropriate soft
matrix element in the given model at the low quark-model scale, where the matrix element is
matched to the QCD result, and 2)~subsequent evolution to higher scales with
appropriate perturbative QCD equations.

For instance, the lattice data correspond typically to the scale of about $%
Q=2$~GeV, as follows from the used value of the lattice spacing,
while the quark model calculation corresponds to a much lower scale,
\begin{equation}
\mu _{0}\sim \Lambda _{\mathrm{QCD}}.
\end{equation}%
A detailed discussion of the evolution issue and ways to set the quark model
scale is presented in Ref.~\cite{Broniowski:2007si,Broniowski:evol}, where the
scale%
\begin{equation}
\mu _{0}=313~\mathrm{MeV}  \label{scale}
\end{equation}%
is advocated. We stress that the inclusion of evolution is crucial for
obtaining the results at experimental or lattice scales. A non-trivial
test is to check that the procedure reproduces consistently other
observables at a given scale, $\mu$ (see e.g.
Ref.~\cite{Broniowski:2007si,Broniowski:evol} for a detailed comparison).

\subsection{Evolution of tGPD \label{sec:evoltGPG}}

The leading-order DGLAP-ERBL evolution for tGPD is given, e.g., in~\cite%
{Belitsky:2005qn}. To carry out this evolution in practical terms, we use
the method given in \cite%
{Kivel:1999sk,Kivel:1999wa,Manashov:2005xp,Kirch:2005tt}, where the basic
objects are the moments in the Gegenbauer polynomials of index $n$
\begin{equation}
g_{n}(\mu)=\int_{0}^{1}dX\, E_{T}^{\pi ,S}(X,\xi ,t;\mu)G_{n}^{3/2}(X/\xi ).
\end{equation}%
The DGLAP region, $X>\xi $, is outside of the orthogonality range for the polynomials $G_{n}^{3/2}(X/\xi)$.
The LO DGLAP-ERBL evolution amounts to the multiplication
\begin{equation}
g_{n}(\mu )=L_{n}g_{n}(\mu _{0}),
\end{equation}%
\begin{equation}
L_{n}=\left( \frac{\alpha (\mu )}{\alpha (\mu _{0})}\right) ^{\gamma
_{n}^{T}/(2\beta _{0})}.
\end{equation}%
The anomalous dimensions in the transversity (tensor) channel are given by
\begin{equation}
\gamma _{n}^{T}=\frac{32}{3}H_{n}-8,
\end{equation}%
where $H_{n}=\sum_{k=1}^{n}1/k$. In particular, one has for the two lowest form factors
$\gamma _{1}^{T}=\frac{8}{3}$ and $\gamma _{2}^{T}=8$. We use $\beta _{0}=%
\frac{11}{3}N_{c}-\frac{2}{3}N_{f}$ and the running coupling constant
\begin{equation}
\alpha (\mu )={4\pi }/[{\beta _{0}\log (\mu ^{2}/\Lambda _{QCD}^{2})}],
\end{equation}%
with $\Lambda _{\mathrm{QCD}}=226~\mathrm{MeV}$ for $N_{c}=N_{f}=3$. The
inversion of the evolved moments back into the evolved GPD, applied in our
calculation, is explained in \cite%
{Kivel:1999sk,Kivel:1999wa,Manashov:2005xp,Kirch:2005tt}.

We also recall that in the transversity channel the quark distributions
evolve autonomously, i.e. do not mix with the gluon distributions, which is
the case of the vector and axial channels. That way no gluon tGPDs are
generated by the QCD evolution, as by construction they vanish in chiral quark
models at the quark-model scale.

\subsection{Evolution of transversity form factors \label{sec:evoltFF}}

The LO DGLAP-ERBL evolution of tGFFs, defined as moments of the GPDs, has
been spelled out explicitly in~\cite{Broniowski:2009zh}. The triangular
structure which appears from the considerations on the evolution of the
tGPDs is, for odd $n=2k+1$,
\begin{eqnarray}
&&B_{2k+1,2l}=k\Gamma (2k)\sum_{m=0}^{k}(4m+3)L_{2m+1}\sum_{j=k-l}^{k} \\
&&\frac{2^{2(j-k)}(-1)^{m-j}\Gamma \left( j+m+\frac{3}{2}\right)
B_{2j+1,2(j-k+l)}^{0}}{\Gamma (2j+1)\Gamma (m-j+1)\Gamma (k-m+1)\Gamma
\left( k+m+\frac{5}{2}\right) },  \notag
\end{eqnarray}%
and, for even $n=2k+2$,
\begin{eqnarray}
&&B_{2k+2,2l}=\Gamma (2k+2)\sum_{m=0}^{k}(4m+5)L_{2m+2}\sum_{j=k-l}^{k} \\
&&\frac{2^{2j-2k-1}(-1)^{m-j}\Gamma \left( j+m+\frac{5}{2}\right)
B_{2(j+1),2(j-k+l)}^{0}}{\Gamma (2j+2)\Gamma (m-j+1)\Gamma (k-m+1)\Gamma
\left( k+m+\frac{7}{2}\right) },  \notag
\end{eqnarray}%
where $k=0,1,2,\dots $ and $l=0,1,\dots ,k$. We have introduced a
short-hand notation $B_{ni}=B_{Tni}^{\pi }(t;\mu )$ and $B_{ni}^{0}=B_{Tni}^{%
\pi }(t;\mu _{0})$. For the lowest moments we have, explicitly,
\begin{eqnarray}
B_{10} &=&L_{1}B_{10}^{0},  \notag \\
B_{32} &=&\frac{1}{5}(L_{1}-L_{3})B_{10}^{0}+L_{3}B_{32}^{0},  \notag \\
B_{54} &=&\frac{1}{105}(9L_{1}-14L_{3}+5L_{5})B_{10}^{0}  \notag \\
&&+\frac{2}{3}(L_{3}-L_{5})B_{32}^{0}+L_{5}B_{54}^{0},  \notag \\
&\dots &  \notag \\
B_{20} &=&L_{2}B_{20}^{0},  \notag \\
B_{42} &=&\frac{3}{7}(L_{2}-L_{4})B_{20}^{0}+L_{4}B_{42}^{0},  \notag \\
&\dots &  \notag \\
B_{30} &=&L_{3}B_{30}^{0},  \notag \\
B_{52} &=&\frac{2}{3}(L_{3}-L_{5})B_{30}^{0}+L_{5}B_{52}^{0},  \notag \\
&\dots &  \notag \\
B_{40} &=&L_{4}B_{40}^{0}.  \label{ev:ns}
\end{eqnarray}%

In particular, the two lowest tGFFs available from the lattice data, $B_{T10}^{\pi
,u}$ and $B_{T20}^{\pi ,u}$, evolve multiplicatively as follows:
\begin{equation*}
B_{Tn0}^{\pi ,u}(t;\mu )=B_{Tn0}^{\pi ,u}(t;\mu _{0})\left( \frac{\alpha
(\mu )}{\alpha (\mu _{0})}\right) ^{\gamma _{n}^{T}/(2\beta _{0})},
\end{equation*}%
which numerically gives
\begin{align}
& B_{T10}^{\pi ,u}(t;2~\mathrm{GeV})=0.75B_{T10}^{\pi ,u}(t;\mu _{0}),
\notag \\
& B_{T20}^{\pi ,u}(t;2~\mathrm{GeV})=0.43B_{T20}^{\pi ,u}(t;\mu _{0}).
\label{evol:explicit}
\end{align}%
Note a stronger reduction for $B_{T20}$ compared to $B_{T10}$ as the result
of the evolution.

In the chiral limit and at $t=0$
\begin{align}
& B_{T10}^{\pi,u}(t=0;\mu_{0})/m_{\pi}=\frac{N_{c} M}{4\pi^{2} f_{\pi}^{2}},
\label{LocLim1} \\
& \frac{B_{T20}^{\pi,u}(t=0;\mu)}{B_{T10}^{\pi,u}(t=0;\mu)}=\frac{1}{3}
\left( \frac{\alpha(\mu)}{\alpha(\mu_{0})}\right) ^{8/27}.  \label{LocLim2}
\end{align}

\section{Numerical results after the QCD evolution \label{sec:res}}

In this section we present our numerical results {\em after the QCD evolution} for the tGPD of the pion, its
special cases $\xi =0$ and $\xi =1$, corresponding to the tPDF and tDA,
respectively, as well as discuss the tGFFs. The latter are compared to the
available lattice data of~\cite{Brommel:2007xd}.

\subsection{tGPD \label{ssec:tgpd}}

\begin{figure}[tb]
\includegraphics[width=.47\textwidth]{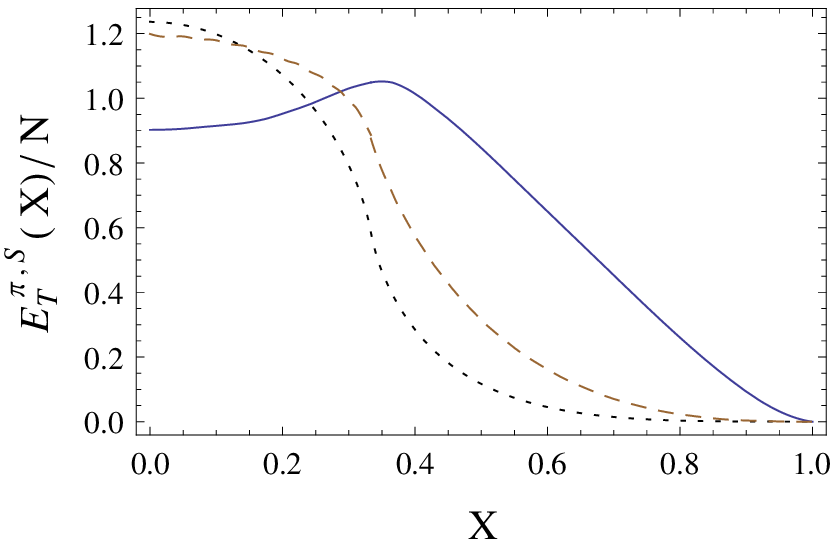} \newline
\vspace{4.5mm} \includegraphics[width=.47\textwidth]{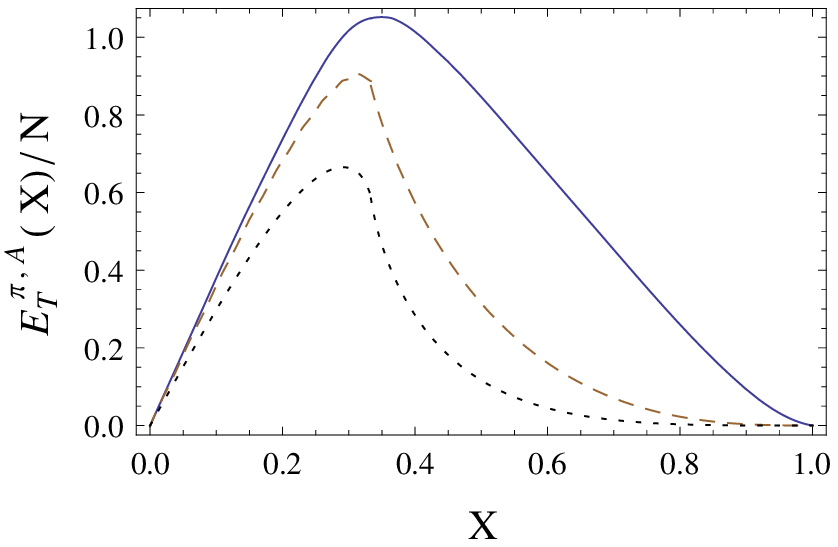}
\caption{(Color online) The DGLAP-ERBL evolution of the symmetric (S, or $I=1$) and
antisymmetric (A, or $I=0$) parts of the quark tGPD of the pion in the non-local HTV
model for $m_\protect\pi=0$, $t=0$,  $\protect\xi=1/3$, and $M=240$~MeV.
The solid line corresponds to the initial condition at the quark model scale
$\protect\mu_0=313$~MeV, the dashed line shows the result of the evolution
to $\protect\mu=2$~GeV, and the dotted line to $\protect\mu=1$~TeV.
\label{fig:tGPDnonloc}}
\end{figure}

\begin{figure}[tb]
\includegraphics[width=.47\textwidth]{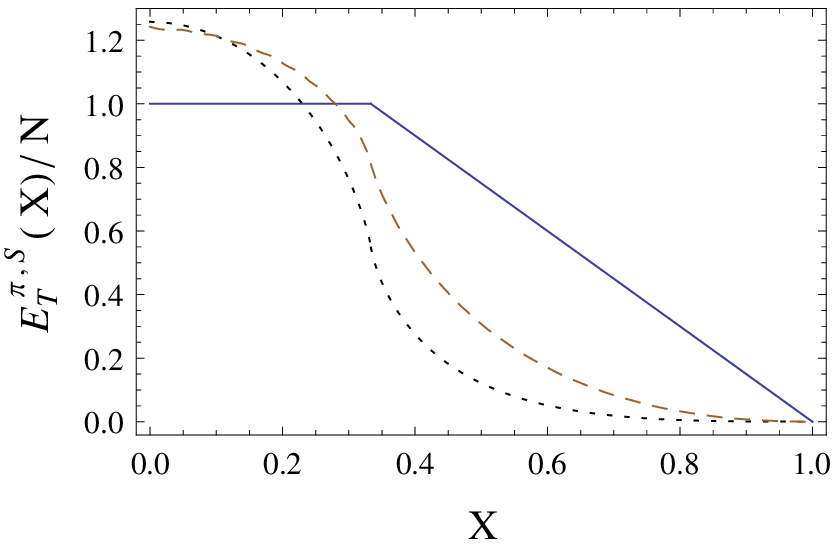}
\newline
\vspace{4.5mm} \includegraphics[width=.47\textwidth]{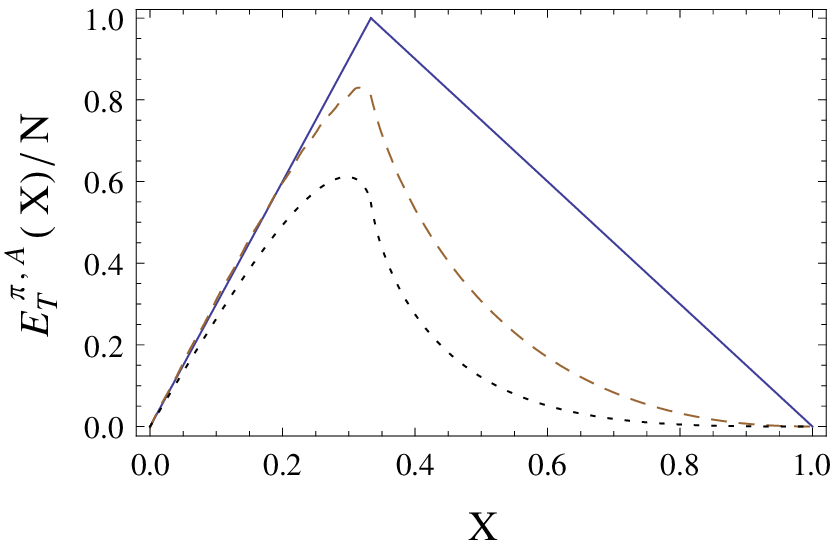}
\caption{(Color online) Same as Fig.~\protect\ref{fig:tGPDnonloc} for the
local NJL model. \label{fig:tGPDloc}}
\end{figure}

The results of the calculation of the tGPD of the pion at a sample value of $%
\xi =1/3$ and at $t=0$, together with the LO DGLAP-ERBL evolution, are given
in Figs.~\ref{fig:tGPDnonloc} and \ref{fig:tGPDloc}. For the non-local case
we take the HTV model (\ref{QPiVertT},\ref{IT}), as the results of the
instanton model (\ref{QPiVertI},\ref{II}) are qualitatively similar. Here we
take for simplicity the chiral limit, $m_{\pi }=0$. We provide in the
figures the symmetric (S) and asymmetric (A) combinations in the $X$
variable (\ref{ETnI01}). The solid lines correspond to the calculation at
the quark-model scale, $\mu _{0}$. In this case we conventionally normalize
the plotted functions with a constant $N$ in such a way that
\begin{equation}
\int_{0}^{1}dXE_{T}^{\pi ,S}(X,\xi ,t=0;\mu _{0})/N=\frac{1+\xi }{2}
\end{equation}%
for all displayed models.

Further, we note the gross qualitative similarity between the nonlocal HTV
model and the local NJL model. The differences are manifest in the end-point
behavior. Near $X=1$ the tGPD in non-local model is suppressed, as explained
in Sect.~\ref{sec:NonLocRes}. Also, near $X=0$ the quantity $E^{\pi,S}_T$ is
depleted compared to the local case, where no minimum is present.

The dashed and dotted curves show the results evolved to the scales $2$~GeV
and 1~TeV, respectively. After the evolution the results of the HTV model
and the local NJL model are qualitatively very similar.

\subsection{tPDF \label{sec:tpdf}}

\begin{figure}[tb]
\includegraphics[width=.47\textwidth]{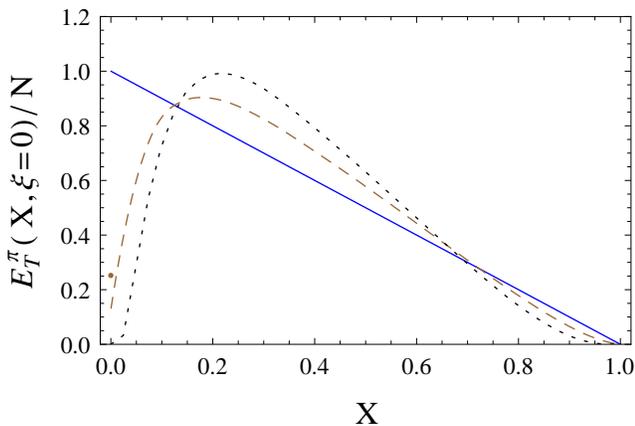}
\caption{(Color online) Comparison of the tPDF ($E_{T}^{\protect\pi }(X,t=0,%
\protect\xi =0)$) in the local model (solid line), instanton model (dashed
line), and the HTV model (dotted line) for $m_{\protect\pi }=0$, evaluated
at the quark-model scale.
\label{fig:tPDF}}
\end{figure}

Next, we explore the special case $\xi =0$, again for $t=0$ and $m_{\pi }=0$%
. In this case tGPD corresponds, by definition, to tPDF. In Fig.~\ref{fig:tPDF} we compare
the predictions of the three considered models at the quark-model scale, $%
\mu _{0}$. We note different end-point behavior, both at $X=1$ and at $X=0$,
according to the discussion presented in Sect.~\ref{sec:NonLocRes}. Near $X=1$
the instanton model has a stronger suppression in tPDF than the HTV model.
The local model approaches zero linearly. Again, we note that the QCD evolution changes the
end-point behavior.

\subsection{tDA \label{sec:tda}}

Another interesting limiting case is provided with \mbox{$\xi=1$}. In that case
\begin{eqnarray}
E_T^\pi(X,t=0,\xi=1)=\phi_T(X),
\end{eqnarray}
where $\phi_T^\pi(X)$ is the tensor distribution amplitude of the pion,
defined as
\begin{eqnarray}
&&\langle 0 | \overline{d}(z) \sigma_{\alpha \beta} \gamma_5 u(-z)| \pi^+(q)
\rangle =  \label{PT} \\
&& i \frac{\sqrt{2}}{3} N^{T} (p_\alpha z_\beta-p_\beta z_\alpha)\int_0^1 du\,
e^{i (2u-1) q\cdot z} \phi_{T}^{\pi}(u),  \notag
\end{eqnarray}
where $X=2u-1$ and $N^{T}$ is the normalization factor yielding $\int_0^1 du
\phi_T(u)=1$.

\begin{figure}[tb]
\includegraphics[width=.47\textwidth]{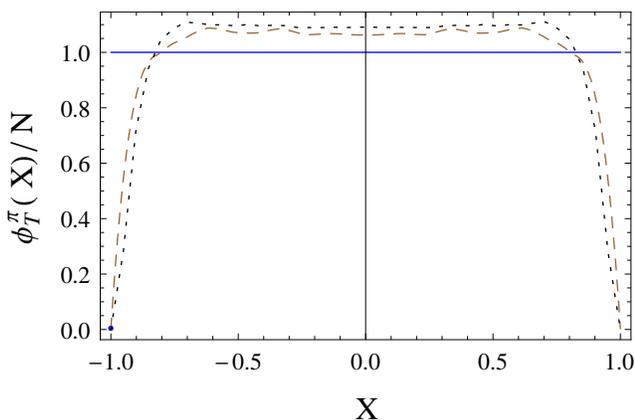}
\caption{(Color online) Comparison of the tDA ($\protect\phi _{T}^{\protect%
\pi }(X)$) in the local model (solid line), instanton model (dashed line),
and the HTV model (dotted line) for $m_{\protect\pi }=0$, evaluated at the
quark-model scale.
\label{fig:tDA}}
\end{figure}

The local NJL model predicts a constant $\phi_T^\pi(X)$ at the quark-model
scale. Again, as seen from Fig.~\ref{fig:tDA}, the difference between the
local and non-local models is seen in the end-point behavior, $X \sim \pm 1$%
. In the intermediate range of $X$ the tDA $\phi_T^\pi(X)$ is close to a
constant also for the non-local models.

\begin{figure}[tb]
\includegraphics[width=.47\textwidth]{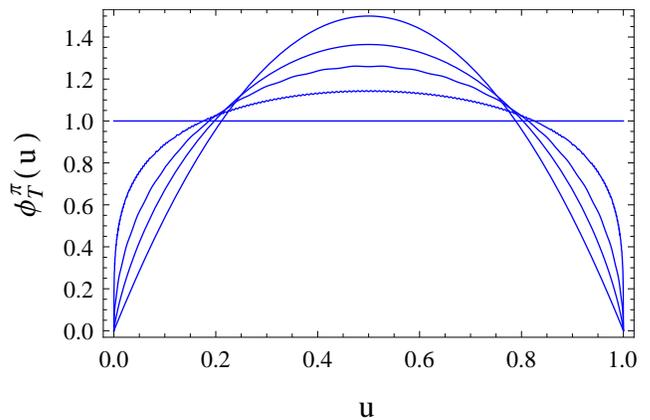}
\caption{(Color online) Evolution of the tensor distribution amplitude, tDA,
in the local NJL model. The subsequent curves (from bottom to top at $u=1/2$) correspond to $\protect\mu =%
\protect\mu _{0}=313$~MeV (the constant), $\protect\mu =500$~MeV, $\protect%
\mu =2$~GeV, $\protect\mu =1000$~GeV, and $\protect\mu =\infty $ (the
asymptotic form $6u(1-u)$.
\label{fig:evphiT}}
\end{figure}

In Fig.~\ref{fig:evphiT} we show the LO ERBL evolution of the tDA of the
pion in the local NJL model. We note a gradual approach towards the
asymptotic form
\begin{eqnarray}
\phi _{T,\mathrm{asym}}^{\pi }(u)=6u(1-u).
\end{eqnarray}
For the non-local models the effect of the evolution is similar.

\subsection{tGFFs \label{ssec:tff}}

\begin{figure}[tb]
\includegraphics[width=.47\textwidth]{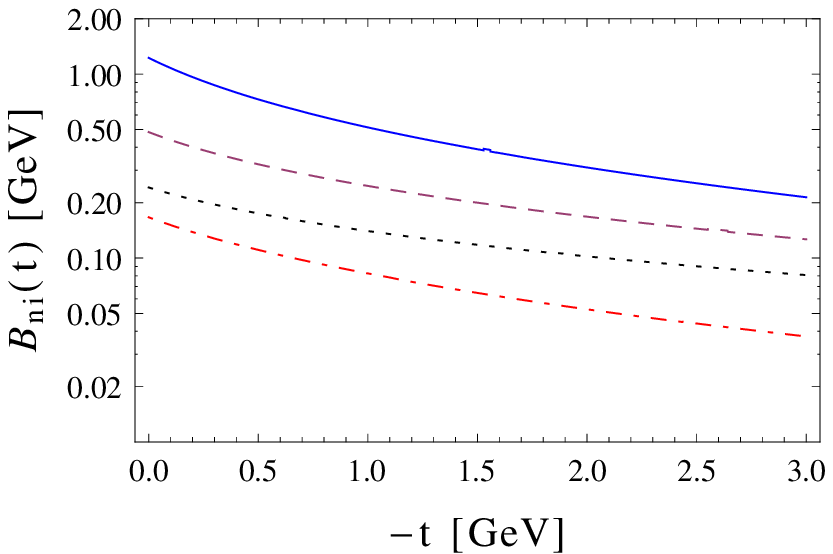} \newline
\vspace{4mm} \includegraphics[width=.47\textwidth]{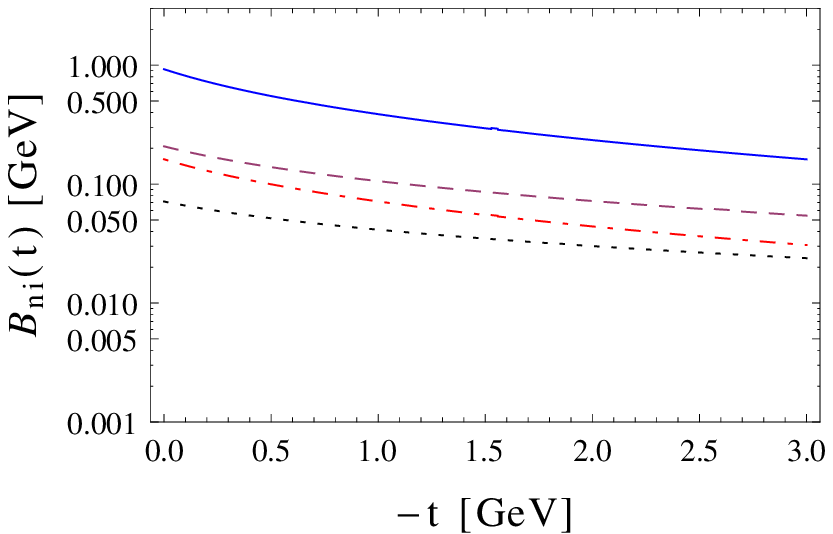}
\caption{(Color online) The transversity form factors $B_{ni}^{u}(t)$,
evaluated in the local NJL model at the quark-model scale $\protect\mu _{0}$
(top panel) and evolved to $\protect\mu =2$~GeV (bottom panel). Solid line
-- $B_{10}^{u}(t)$, dashed line -- $B_{20}^{u}(t)$, dotted line -- $%
B_{30}^{u}(t)$, dash-dotted line -- $B_{32}^{u}(t)$.
\label{fig:evff}}
\end{figure}

In Fig.~\ref{fig:evff} we show the LO DGLAP-ERBL evolution of the tGFFs
evaluated in the local NJL model.

By comparing the two panels of Fig.~\ref{fig:evff} we note that for the tGFFs
is multiplicative, and increasing the scale leads a quenching of $B_{Tn0}$
the form factor. For the form factors $B_{Tni}$ with $i\neq 0$ the evolution
is more complicated, as can be inferred from Eq.~(\ref{ev:ns}).
For the non-local models the effects of the evolution for tGFFs are similar.

\subsection{Chiral quark models vs lattice \label{sec:lattice}}

The content of this Section has already been presented by us in a greater
detail in~\cite{Broniowski:2010nt}. For the completeness of the present work
we repeat the main results.

The presently available full-QCD lattice results \cite{Brommel:2007xd} are
for $B_{10}^{\pi ,u}$ and $B_{20}^{\pi ,u}$ and for $-t$ up to 2.5~GeV$^{2}$%
, with moderately low, but still away from the physical limit, values of the
pion mass, $m_{\pi }\sim 600$~MeV. The calculation of~\cite{Brommel:2007xd}
uses the same $N_{f}=2$ set of the QCDSF/UKQCD ensembles with improved
Wilson fermions and the Wilson gauge action that were used previously in the
analysis of the pion charge and gravitational form factors \cite%
{Brommel:2005ee}.

We note that for $t=0$ both the local and non-local models yield the
normalization
\begin{eqnarray}
& B_{T10}^{\pi,u}(t=0;\mu_{0})/m_{\pi}=\frac{N_{c}}{2\pi^{2}f_{\pi}^{2}}
\notag \\
& \times\int_{0}^{\infty}du\frac{um^{2}(u)}{(u+m^{2}(u))^{3}}%
(m(u)-um^{\prime }(u)),  \label{NonLocB1} \\
& B_{T20}^{\pi,u}(t=0;\mu_{0})/m_{\pi}=\frac{N_{c}}{2\pi^{2}f_{\pi}^{2}}%
\Big\{\int_{0}^{\infty}du\frac{um(u)}{(u+m^{2}(u))^{3}}  \notag \\
& \times(m^{2}(u)+\frac{1}{2}um(u)m^{\prime}(u)+\frac{1}{6}%
u^{2}m^{\prime,2}(u))  \notag \\
& -\int_{0}^{\infty}du\frac{u^{2}m^{2}(u)}{(u+m^{2}(u))^{4}}
(m(u)+2m^{2}(u)m^{\prime}(u))\Big\},  \label{NonLocB2}
\end{eqnarray}
where $m^{\prime}(u)=dm(u)/du$. In the local limit, where $m(k^{2})\to%
\mathrm{const}$, one reproduces Eqs.~(\ref{LocLim1},\ref{LocLim2}).

\begin{figure}[tb]
\includegraphics[width=.47\textwidth]{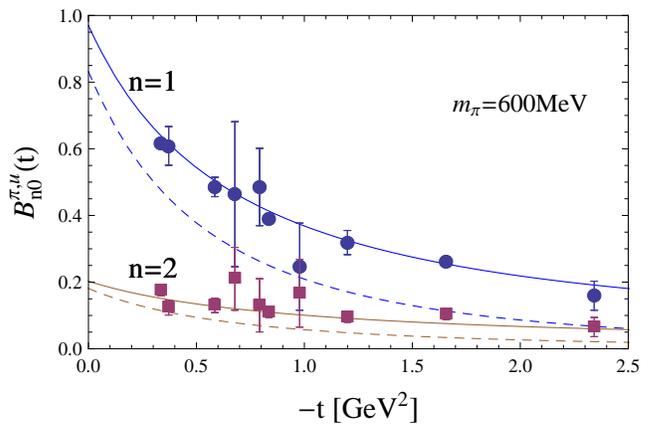} \vspace{-2mm}
\caption{(Color online) The transversity form factors in the HTV model
(solid line) and in the instanton model (dashed line). The data
come from~\protect\cite{Brommel:2005ee}.
\label{fig:nonloc}}
\end{figure}

The results for $B_{Tn0}^{\pi,u}(t)$, $n=1,2$, are shown in Fig.~\ref%
{fig:nonloc}. In our study we have assumed that $B_{Tn0}/m_{\pi}$ depends
weakly on $m_{\pi}$, similarly to the local model case~\cite%
{Broniowski:2010nt}. Therefore, to compare to the lattice data for $B_{Tn0}$,
we multiply the results of the calculations obtained in the chiral limit
with $m_{\pi}=600$~MeV. We have carried out the QCD evolution procedure as
described in the previous Sections, from the quark model scale up to the
lattice scale of 2~GeV. From Fig.~\ref{fig:nonloc} we note that the HTV
model with the vertex function given by Eq.~(\ref{QPiVertT}) (solid lines)
and with $M_q=300$~MeV works best, describing accurately the data, while
the instanton model, Eq.~(\ref{QPiVertI}) (dashed lines), results in form
factors falling-off too steeply. We have found that lower values of $M_q$
spoil the agreement with the lattice data.

\begin{figure}[tb]
\includegraphics[width=.47\textwidth]{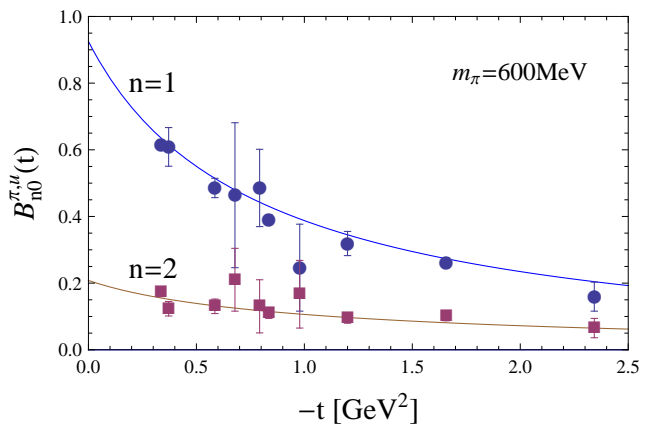} \vspace{-2mm}
\caption{(Color online) The transversity form factors obtained in the NJL
model (lines) for $M=250$~MeV and $m_{}\protect\pi=600$~MeV, evolved to the
lattice scale of 2~GeV and compared to the lattice data from Fig.~1 of~%
\protect\cite{Brommel:2007xd} (points).
\label{fig:resu}}
\end{figure}

In Fig.~\ref{fig:resu} we show the results from the local NJL model evolved
to the lattice scale of $\mu=2$~GeV, confronted with the lattice data
scanned from Fig.~1 of~\cite{Brommel:2007xd}. We have used $m_{\pi}=600$~MeV
and selected $M=250$~MeV, which optimizes the comparison. As we see, the
agreement is remarkable.

In Ref.~\cite{Broniowski:2010nt} we have also investigated the dependence of
the values of the form factors at $t=0$ on the value of $m_{\pi}$, as
studied in~\cite{Brommel:2007xd}.
We have also noted in \cite{Broniowski:2010nt} that the results presented in Fig.~%
\ref{fig:resu} depend quite sensitively on the value of the constituent
quark mass, $M$, with higher $M$ yielding lower values of the transversity
form factors.

\section{Conclusions \label{sec:concl}}

In the present paper we have shown how the spinless pion acquires a
non-trivial spin structure within the framework of chiral quark
models. This has been achieved by computing the transversity
distributions, corresponding to matrix elements of the tensor quark
density, within chiral quark models, where the pion arises as the
pseudo-Goldstone boson of the spontaneously broken chiral
symmetry. Moreover, we have worked at the leading order in the $1/N_c$
expansion, which amounts to carrying out one-quark-loop calculations,
where the implementation of the symmetry constraints becomes
absolutely essential. Chiral symmetry is respected by implementing the
pertinent chiral Ward-Takahashi identities at the quark
level. Moreover, the relativity constraints are fulfilled in terms of
the polynomiality conditions which are manifestly preserved through
the use of the double distributions, or, equivalently, by working with
the $\alpha$-representations.

We have provided comprehensive results for the tGPDs of the pion, as well as
related quantities following from restrained kinematics, evaluation of
moments, or taking the Fourier-Bessel transforms to the
impact-parameter space. We have also shown in detail various technical
aspects of our analysis, including the use of the
$\alpha$-representation in the nonlocal models.

The generated tGPDs are defined at
a given low-energy quark-model scale, and comparison to data or lattice results
corresponds to implementing the suitable QCD evolution. Actually, while the
momentum-transfer or, equivalently, the impact-parameter dependence of
the tGFFs remains scale independent, their absolute normalization does
depend multiplicatively on the renormalization scale.  Remarkably, the
absolute predictions for the multiplicatively evolved $B_{Tn0}$, for $n=1,2$, agree
surprisingly well with the lattice results, supporting many previous calculations
following the same chiral-quark-model scheme amended with the subsequent QCD evolution.

\bigskip

One of us (AED) is thankful A.V. Radyushkin and S.V. Mikhailov for numerous discussions.

\medskip

{Supported by the Bogoliubov-Infeld program (JINR), the Polish Ministry of
Science and Higher Education, grants N~N202~263438 and N~N202~249235,
Spanish DGI and FEDER grant FIS2008-01143/FIS, Junta de Andaluc{\'{\i}}a
grant FQM225-05, and EU Integrated Infrastructure Initiative Hadron Physics
Project, contract RII3-CT-2004-506078. AED acknowledges partial support from
the Russian Foundation for Basic Research, projects No.~10-02-00368 and No.~11-02-00112.}


\begin{thebibliography}{96}
\expandafter\ifx\csname natexlab\endcsname\relax\def\natexlab#1{#1}\fi
\expandafter\ifx\csname bibnamefont\endcsname\relax
  \def\bibnamefont#1{#1}\fi
\expandafter\ifx\csname bibfnamefont\endcsname\relax
  \def\bibfnamefont#1{#1}\fi
\expandafter\ifx\csname citenamefont\endcsname\relax
  \def\citenamefont#1{#1}\fi
\expandafter\ifx\csname url\endcsname\relax
  \def\url#1{\texttt{#1}}\fi
\expandafter\ifx\csname urlprefix\endcsname\relax\def\urlprefix{URL }\fi
\providecommand{\bibinfo}[2]{#2}
\providecommand{\eprint}[2][]{\url{#2}}

\bibitem[{\citenamefont{Callan and Gross}(1969)}]{Callan:1969uq}
\bibinfo{author}{\bibfnamefont{C.~G.} \bibnamefont{Callan}, \bibfnamefont{Jr.}}
  \bibnamefont{and} \bibinfo{author}{\bibfnamefont{D.~J.} \bibnamefont{Gross}},
  \bibinfo{journal}{Phys. Rev. Lett.} \textbf{\bibinfo{volume}{22}},
  \bibinfo{pages}{156} (\bibinfo{year}{1969}).

\bibitem[{\citenamefont{Barone et~al.}(2002)\citenamefont{Barone, Drago, and
  Ratcliffe}}]{Barone:2001sp}
\bibinfo{author}{\bibfnamefont{V.}~\bibnamefont{Barone}},
  \bibinfo{author}{\bibfnamefont{A.}~\bibnamefont{Drago}}, \bibnamefont{and}
  \bibinfo{author}{\bibfnamefont{P.~G.} \bibnamefont{Ratcliffe}},
  \bibinfo{journal}{Phys. Rept.} \textbf{\bibinfo{volume}{359}},
  \bibinfo{pages}{1} (\bibinfo{year}{2002}), \eprint{hep-ph/0104283}.

\bibitem[{\citenamefont{Mueller et~al.}(1994)\citenamefont{Mueller, Robaschik,
  Geyer, Dittes, and Horejsi}}]{Mueller:1998fv}
\bibinfo{author}{\bibfnamefont{D.}~\bibnamefont{Mueller}},
  \bibinfo{author}{\bibfnamefont{D.}~\bibnamefont{Robaschik}},
  \bibinfo{author}{\bibfnamefont{B.}~\bibnamefont{Geyer}},
  \bibinfo{author}{\bibfnamefont{F.~M.} \bibnamefont{Dittes}},
  \bibnamefont{and} \bibinfo{author}{\bibfnamefont{J.}~\bibnamefont{Horejsi}},
  \bibinfo{journal}{Fortschr. Phys.} \textbf{\bibinfo{volume}{42}},
  \bibinfo{pages}{101} (\bibinfo{year}{1994}), \eprint{hep-ph/9812448}.

\bibitem[{\citenamefont{Ji}(1997)}]{Ji:1996ek}
\bibinfo{author}{\bibfnamefont{X.-D.} \bibnamefont{Ji}},
  \bibinfo{journal}{Phys. Rev. Lett.} \textbf{\bibinfo{volume}{78}},
  \bibinfo{pages}{610} (\bibinfo{year}{1997}), \eprint{hep-ph/9603249}.

\bibitem[{\citenamefont{Radyushkin}(1996)}]{Radyushkin:1996nd}
\bibinfo{author}{\bibfnamefont{A.~V.} \bibnamefont{Radyushkin}},
  \bibinfo{journal}{Phys. Lett.} \textbf{\bibinfo{volume}{B380}},
  \bibinfo{pages}{417} (\bibinfo{year}{1996}), \eprint{hep-ph/9604317}.

\bibitem[{\citenamefont{Belitsky and Radyushkin}(2005)}]{Belitsky:2005qn}
\bibinfo{author}{\bibfnamefont{A.~V.} \bibnamefont{Belitsky}} \bibnamefont{and}
  \bibinfo{author}{\bibfnamefont{A.~V.} \bibnamefont{Radyushkin}},
  \bibinfo{journal}{Phys. Rept.} \textbf{\bibinfo{volume}{418}},
  \bibinfo{pages}{1} (\bibinfo{year}{2005}), \eprint{hep-ph/0504030}.

\bibitem[{\citenamefont{Feldmann}(2007)}]{Feldmann:2007zz}
\bibinfo{author}{\bibfnamefont{T.}~\bibnamefont{Feldmann}},
  \bibinfo{journal}{Eur. Phys. J. Special Topics}
  \textbf{\bibinfo{volume}{140}}, \bibinfo{pages}{135} (\bibinfo{year}{2007}).

\bibitem[{\citenamefont{Boffi and Pasquini}(2007)}]{Boffi:2007yc}
\bibinfo{author}{\bibfnamefont{S.}~\bibnamefont{Boffi}} \bibnamefont{and}
  \bibinfo{author}{\bibfnamefont{B.}~\bibnamefont{Pasquini}},
  \bibinfo{journal}{Riv. Nuovo Cim.} \textbf{\bibinfo{volume}{30}},
  \bibinfo{pages}{387} (\bibinfo{year}{2007}), \eprint{0711.2625}.

\bibitem[{\citenamefont{Burkardt}(2000)}]{Burkardt:2000za}
\bibinfo{author}{\bibfnamefont{M.}~\bibnamefont{Burkardt}},
  \bibinfo{journal}{Phys. Rev.} \textbf{\bibinfo{volume}{D62}},
  \bibinfo{pages}{071503} (\bibinfo{year}{2000}), \eprint{hep-ph/0005108}.

\bibitem[{\citenamefont{Burkardt}(2003)}]{Burkardt:2002hr}
\bibinfo{author}{\bibfnamefont{M.}~\bibnamefont{Burkardt}},
  \bibinfo{journal}{Int. J. Mod. Phys.} \textbf{\bibinfo{volume}{A18}},
  \bibinfo{pages}{173} (\bibinfo{year}{2003}), \eprint{hep-ph/0207047}.

\bibitem[{\citenamefont{Burkardt and Dalley}(2002)}]{Burkardt:2001jg}
\bibinfo{author}{\bibfnamefont{M.}~\bibnamefont{Burkardt}} \bibnamefont{and}
  \bibinfo{author}{\bibfnamefont{S.}~\bibnamefont{Dalley}},
  \bibinfo{journal}{Prog. Part. Nucl. Phys.} \textbf{\bibinfo{volume}{48}},
  \bibinfo{pages}{317} (\bibinfo{year}{2002}), \eprint{hep-ph/0112007}.

\bibitem[{\citenamefont{Musch et~al.}(2011)\citenamefont{Musch, Hagler, Negele,
  and Schafer}}]{Musch:2010ka}
\bibinfo{author}{\bibfnamefont{B.~U.} \bibnamefont{Musch}},
  \bibinfo{author}{\bibfnamefont{P.}~\bibnamefont{Hagler}},
  \bibinfo{author}{\bibfnamefont{J.~W.} \bibnamefont{Negele}},
  \bibnamefont{and} \bibinfo{author}{\bibfnamefont{A.}~\bibnamefont{Schafer}},
  \bibinfo{journal}{Phys. Rev.} \textbf{\bibinfo{volume}{D83}},
  \bibinfo{pages}{094507} (\bibinfo{year}{2011}), \eprint{1011.1213}.

\bibitem[{\citenamefont{Hagler}(2010)}]{Hagler:2009ni}
\bibinfo{author}{\bibfnamefont{P.}~\bibnamefont{Hagler}},
  \bibinfo{journal}{Phys. Rept.} \textbf{\bibinfo{volume}{490}},
  \bibinfo{pages}{49} (\bibinfo{year}{2010}), \eprint{0912.5483}.

\bibitem[{\citenamefont{Broniowski and Arriola}(2009)}]{Broniowski:2009zh}
\bibinfo{author}{\bibfnamefont{W.}~\bibnamefont{Broniowski}} \bibnamefont{and}
  \bibinfo{author}{\bibfnamefont{E.~R.} \bibnamefont{Arriola}},
  \bibinfo{journal}{Phys. Rev.} \textbf{\bibinfo{volume}{D79}},
  \bibinfo{pages}{057501} (\bibinfo{year}{2009}), \eprint{0901.3336}.

\bibitem[{\citenamefont{Diehl and Hagler}(2005)}]{Diehl:2005jf}
\bibinfo{author}{\bibfnamefont{M.}~\bibnamefont{Diehl}} \bibnamefont{and}
  \bibinfo{author}{\bibfnamefont{P.}~\bibnamefont{Hagler}},
  \bibinfo{journal}{Eur. Phys. J.} \textbf{\bibinfo{volume}{C44}},
  \bibinfo{pages}{87} (\bibinfo{year}{2005}), \eprint{hep-ph/0504175}.

\bibitem[{\citenamefont{Burkardt}(2005)}]{Burkardt:2005hp}
\bibinfo{author}{\bibfnamefont{M.}~\bibnamefont{Burkardt}},
  \bibinfo{journal}{Phys. Rev.} \textbf{\bibinfo{volume}{D72}},
  \bibinfo{pages}{094020} (\bibinfo{year}{2005}), \eprint{hep-ph/0505189}.

\bibitem[{\citenamefont{Brommel et~al.}(2008)}]{Brommel:2007xd}
\bibinfo{author}{\bibfnamefont{D.}~\bibnamefont{Brommel}} \bibnamefont{et~al.}
  (\bibinfo{collaboration}{QCDSF}), \bibinfo{journal}{Phys. Rev. Lett.}
  \textbf{\bibinfo{volume}{101}}, \bibinfo{pages}{122001}
  (\bibinfo{year}{2008}), \eprint{0708.2249}.

\bibitem[{\citenamefont{Diehl and Szymanowski}(2010)}]{Diehl:2010ru}
\bibinfo{author}{\bibfnamefont{M.}~\bibnamefont{Diehl}} \bibnamefont{and}
  \bibinfo{author}{\bibfnamefont{L.}~\bibnamefont{Szymanowski}},
  \bibinfo{journal}{Phys. Lett.} \textbf{\bibinfo{volume}{B690}},
  \bibinfo{pages}{149} (\bibinfo{year}{2010}), \eprint{1003.4171}.

\bibitem[{\citenamefont{Broniowski et~al.}(2010)\citenamefont{Broniowski,
  Dorokhov, and Arriola}}]{Broniowski:2010nt}
\bibinfo{author}{\bibfnamefont{W.}~\bibnamefont{Broniowski}},
  \bibinfo{author}{\bibfnamefont{A.~E.} \bibnamefont{Dorokhov}},
  \bibnamefont{and} \bibinfo{author}{\bibfnamefont{E.~R.}
  \bibnamefont{Arriola}}, \bibinfo{journal}{Phys. Rev.}
  \textbf{\bibinfo{volume}{D82}}, \bibinfo{pages}{094001}
  (\bibinfo{year}{2010}), \eprint{1007.4960}.

\bibitem[{\citenamefont{Nam and Kim}(2011)}]{Nam:2010pt}
\bibinfo{author}{\bibfnamefont{S.-i.} \bibnamefont{Nam}} \bibnamefont{and}
  \bibinfo{author}{\bibfnamefont{H.-C.} \bibnamefont{Kim}},
  \bibinfo{journal}{Phys. Lett.} \textbf{\bibinfo{volume}{B700}},
  \bibinfo{pages}{305} (\bibinfo{year}{2011}), \eprint{1010.0468}.

\bibitem[{\citenamefont{Ji}(1998)}]{Ji:1998pc}
\bibinfo{author}{\bibfnamefont{X.-D.} \bibnamefont{Ji}}, \bibinfo{journal}{J.
  Phys.} \textbf{\bibinfo{volume}{G24}}, \bibinfo{pages}{1181}
  (\bibinfo{year}{1998}), \eprint{hep-ph/9807358}.

\bibitem[{\citenamefont{Radyushkin}(2002)}]{Radyushkin:2000uy}
\bibinfo{author}{\bibfnamefont{A.~V.} \bibnamefont{Radyushkin}}, in
  \emph{\bibinfo{booktitle}{At the frontier of particle physics. Handbook of
  QCD}}, edited by \bibinfo{editor}{\bibfnamefont{M.}~\bibnamefont{Shifman}}
  (\bibinfo{publisher}{World Scientific}, \bibinfo{address}{Singapore},
  \bibinfo{year}{2002}), \eprint{hep-ph/0101225}.

\bibitem[{\citenamefont{Goeke et~al.}(2001)\citenamefont{Goeke, Polyakov, and
  Vanderhaeghen}}]{Goeke:2001tz}
\bibinfo{author}{\bibfnamefont{K.}~\bibnamefont{Goeke}},
  \bibinfo{author}{\bibfnamefont{M.~V.} \bibnamefont{Polyakov}},
  \bibnamefont{and}
  \bibinfo{author}{\bibfnamefont{M.}~\bibnamefont{Vanderhaeghen}},
  \bibinfo{journal}{Prog. Part. Nucl. Phys.} \textbf{\bibinfo{volume}{47}},
  \bibinfo{pages}{401} (\bibinfo{year}{2001}), \eprint{hep-ph/0106012}.

\bibitem[{\citenamefont{Bakulev et~al.}(2000)\citenamefont{Bakulev, Ruskov,
  Goeke, and Stefanis}}]{Bakulev:2000eb}
\bibinfo{author}{\bibfnamefont{A.~P.} \bibnamefont{Bakulev}},
  \bibinfo{author}{\bibfnamefont{R.}~\bibnamefont{Ruskov}},
  \bibinfo{author}{\bibfnamefont{K.}~\bibnamefont{Goeke}}, \bibnamefont{and}
  \bibinfo{author}{\bibfnamefont{N.~G.} \bibnamefont{Stefanis}},
  \bibinfo{journal}{Phys. Rev.} \textbf{\bibinfo{volume}{D62}},
  \bibinfo{pages}{054018} (\bibinfo{year}{2000}), \eprint{hep-ph/0004111}.

\bibitem[{\citenamefont{Diehl}(2003)}]{Diehl:2003ny}
\bibinfo{author}{\bibfnamefont{M.}~\bibnamefont{Diehl}},
  \bibinfo{journal}{Phys. Rept.} \textbf{\bibinfo{volume}{388}},
  \bibinfo{pages}{41} (\bibinfo{year}{2003}), \eprint{hep-ph/0307382}.

\bibitem[{\citenamefont{Ji}(2004)}]{Ji:2004gf}
\bibinfo{author}{\bibfnamefont{X.-D.} \bibnamefont{Ji}}, \bibinfo{journal}{Ann.
  Rev. Nucl. Part. Sci.} \textbf{\bibinfo{volume}{54}}, \bibinfo{pages}{413}
  (\bibinfo{year}{2004}).

\bibitem[{\citenamefont{Davidson and Ruiz~Arriola}(1995)}]{Davidson:1994uv}
\bibinfo{author}{\bibfnamefont{R.~M.} \bibnamefont{Davidson}} \bibnamefont{and}
  \bibinfo{author}{\bibfnamefont{E.}~\bibnamefont{Ruiz~Arriola}},
  \bibinfo{journal}{Phys. Lett.} \textbf{\bibinfo{volume}{B348}},
  \bibinfo{pages}{163} (\bibinfo{year}{1995}).

\bibitem[{\citenamefont{Ruiz~Arriola}(2001)}]{RuizArriola:2001rr}
\bibinfo{author}{\bibfnamefont{E.}~\bibnamefont{Ruiz~Arriola}}, in
  \emph{\bibinfo{booktitle}{proc. of the workshop Lepton Scattering, Hadrons
  and QCD, Adelaide, Australia, 2001}}, edited by
  \bibinfo{editor}{\bibnamefont{{W. Melnitchouk et al.}}}
  (\bibinfo{publisher}{World Scientific}, \bibinfo{address}{Singapore},
  \bibinfo{year}{2001}), \eprint{hep-ph/0107087}.

\bibitem[{\citenamefont{Davidson and Ruiz~Arriola}(2002)}]{Davidson:2001cc}
\bibinfo{author}{\bibfnamefont{R.~M.} \bibnamefont{Davidson}} \bibnamefont{and}
  \bibinfo{author}{\bibfnamefont{E.}~\bibnamefont{Ruiz~Arriola}},
  \bibinfo{journal}{Acta Phys. Polon.} \textbf{\bibinfo{volume}{B33}},
  \bibinfo{pages}{1791} (\bibinfo{year}{2002}), \eprint{hep-ph/0110291}.

\bibitem[{\citenamefont{Broniowski and Ruiz~Arriola}(2003)}]{Broniowski:2003rp}
\bibinfo{author}{\bibfnamefont{W.}~\bibnamefont{Broniowski}} \bibnamefont{and}
  \bibinfo{author}{\bibfnamefont{E.}~\bibnamefont{Ruiz~Arriola}},
  \bibinfo{journal}{Phys. Lett.} \textbf{\bibinfo{volume}{B574}},
  \bibinfo{pages}{57} (\bibinfo{year}{2003}), \eprint{hep-ph/0307198}.

\bibitem[{\citenamefont{Dorokhov and Tomio}(1998)}]{Dorokhov:1998up}
\bibinfo{author}{\bibfnamefont{A.~E.} \bibnamefont{Dorokhov}} \bibnamefont{and}
  \bibinfo{author}{\bibfnamefont{L.}~\bibnamefont{Tomio}}
  (\bibinfo{year}{1998}), \eprint{hep-ph/9803329}.

\bibitem[{\citenamefont{Polyakov and Weiss}(1999)}]{Polyakov:1999gs}
\bibinfo{author}{\bibfnamefont{M.~V.} \bibnamefont{Polyakov}} \bibnamefont{and}
  \bibinfo{author}{\bibfnamefont{C.}~\bibnamefont{Weiss}},
  \bibinfo{journal}{Phys. Rev.} \textbf{\bibinfo{volume}{D60}},
  \bibinfo{pages}{114017} (\bibinfo{year}{1999}), \eprint{hep-ph/9902451}.

\bibitem[{\citenamefont{Dorokhov and Tomio}(2000)}]{Dorokhov:2000gu}
\bibinfo{author}{\bibfnamefont{A.~E.} \bibnamefont{Dorokhov}} \bibnamefont{and}
  \bibinfo{author}{\bibfnamefont{L.}~\bibnamefont{Tomio}},
  \bibinfo{journal}{Phys. Rev.} \textbf{\bibinfo{volume}{D62}},
  \bibinfo{pages}{014016} (\bibinfo{year}{2000}).

\bibitem[{\citenamefont{Anikin et~al.}(2000{\natexlab{a}})\citenamefont{Anikin,
  Dorokhov, Maksimov, Tomio, and Vento}}]{Anikin:2000th}
\bibinfo{author}{\bibfnamefont{I.~V.} \bibnamefont{Anikin}},
  \bibinfo{author}{\bibfnamefont{A.~E.} \bibnamefont{Dorokhov}},
  \bibinfo{author}{\bibfnamefont{A.~E.} \bibnamefont{Maksimov}},
  \bibinfo{author}{\bibfnamefont{L.}~\bibnamefont{Tomio}}, \bibnamefont{and}
  \bibinfo{author}{\bibfnamefont{V.}~\bibnamefont{Vento}},
  \bibinfo{journal}{Nucl. Phys.} \textbf{\bibinfo{volume}{A678}},
  \bibinfo{pages}{175} (\bibinfo{year}{2000}{\natexlab{a}}).

\bibitem[{\citenamefont{Praszalowicz and
  Rostworowski}(2002)}]{Praszalowicz:2002ct}
\bibinfo{author}{\bibfnamefont{M.}~\bibnamefont{Praszalowicz}}
  \bibnamefont{and}
  \bibinfo{author}{\bibfnamefont{A.}~\bibnamefont{Rostworowski}}, in
  \emph{\bibinfo{booktitle}{proc. of the XXXVIIth Rencontres de Moriond}}
  (\bibinfo{year}{2002}), \eprint{hep-ph/0205177}.

\bibitem[{\citenamefont{Praszalowicz and
  Rostworowski}(2003)}]{Praszalowicz:2003pr}
\bibinfo{author}{\bibfnamefont{M.}~\bibnamefont{Praszalowicz}}
  \bibnamefont{and}
  \bibinfo{author}{\bibfnamefont{A.}~\bibnamefont{Rostworowski}},
  \bibinfo{journal}{Acta Phys. Polon.} \textbf{\bibinfo{volume}{B34}},
  \bibinfo{pages}{2699} (\bibinfo{year}{2003}), \eprint{hep-ph/0302269}.

\bibitem[{\citenamefont{Bzdak and Praszalowicz}(2003)}]{Bzdak:2003qe}
\bibinfo{author}{\bibfnamefont{A.}~\bibnamefont{Bzdak}} \bibnamefont{and}
  \bibinfo{author}{\bibfnamefont{M.}~\bibnamefont{Praszalowicz}},
  \bibinfo{journal}{Acta Phys. Polon.} \textbf{\bibinfo{volume}{B34}},
  \bibinfo{pages}{3401} (\bibinfo{year}{2003}), \eprint{hep-ph/0305217}.

\bibitem[{\citenamefont{Holt and Roberts}(2010)}]{Holt:2010vj}
\bibinfo{author}{\bibfnamefont{R.~J.} \bibnamefont{Holt}} \bibnamefont{and}
  \bibinfo{author}{\bibfnamefont{C.~D.} \bibnamefont{Roberts}},
  \bibinfo{journal}{Rev. Mod. Phys.} \textbf{\bibinfo{volume}{82}},
  \bibinfo{pages}{2991} (\bibinfo{year}{2010}), \eprint{1002.4666}.

\bibitem[{\citenamefont{Nguyen et~al.}(2011)\citenamefont{Nguyen, Bashir,
  Roberts, and Tandy}}]{Nguyen:2011jy}
\bibinfo{author}{\bibfnamefont{T.}~\bibnamefont{Nguyen}},
  \bibinfo{author}{\bibfnamefont{A.}~\bibnamefont{Bashir}},
  \bibinfo{author}{\bibfnamefont{C.~D.} \bibnamefont{Roberts}},
  \bibnamefont{and} \bibinfo{author}{\bibfnamefont{P.~C.} \bibnamefont{Tandy}},
  \bibinfo{journal}{Phys. Rev.} \textbf{\bibinfo{volume}{C83}},
  \bibinfo{pages}{062201} (\bibinfo{year}{2011}), \eprint{1102.2448}.

\bibitem[{\citenamefont{Theussl et~al.}(2004)\citenamefont{Theussl, Noguera,
  and Vento}}]{Theussl:2002xp}
\bibinfo{author}{\bibfnamefont{L.}~\bibnamefont{Theussl}},
  \bibinfo{author}{\bibfnamefont{S.}~\bibnamefont{Noguera}}, \bibnamefont{and}
  \bibinfo{author}{\bibfnamefont{V.}~\bibnamefont{Vento}},
  \bibinfo{journal}{Eur. Phys. J.} \textbf{\bibinfo{volume}{A20}},
  \bibinfo{pages}{483} (\bibinfo{year}{2004}), \eprint{nucl-th/0211036}.

\bibitem[{\citenamefont{Bissey et~al.}(2004)\citenamefont{Bissey, Cudell,
  Cugnon, Lansberg, and Stassart}}]{Bissey:2003yr}
\bibinfo{author}{\bibfnamefont{F.}~\bibnamefont{Bissey}},
  \bibinfo{author}{\bibfnamefont{J.~R.} \bibnamefont{Cudell}},
  \bibinfo{author}{\bibfnamefont{J.}~\bibnamefont{Cugnon}},
  \bibinfo{author}{\bibfnamefont{J.~P.} \bibnamefont{Lansberg}},
  \bibnamefont{and} \bibinfo{author}{\bibfnamefont{P.}~\bibnamefont{Stassart}},
  \bibinfo{journal}{Phys. Lett.} \textbf{\bibinfo{volume}{B587}},
  \bibinfo{pages}{189} (\bibinfo{year}{2004}), \eprint{hep-ph/0310184}.

\bibitem[{\citenamefont{Noguera and Vento}(2006)}]{Noguera:2005cc}
\bibinfo{author}{\bibfnamefont{S.}~\bibnamefont{Noguera}} \bibnamefont{and}
  \bibinfo{author}{\bibfnamefont{V.}~\bibnamefont{Vento}},
  \bibinfo{journal}{Eur. Phys. J.} \textbf{\bibinfo{volume}{A28}},
  \bibinfo{pages}{227} (\bibinfo{year}{2006}), \eprint{hep-ph/0505102}.

\bibitem[{\citenamefont{Broniowski
  et~al.}(2008{\natexlab{a}})\citenamefont{Broniowski, Ruiz~Arriola, and
  Golec-Biernat}}]{Broniowski:2007si}
\bibinfo{author}{\bibfnamefont{W.}~\bibnamefont{Broniowski}},
  \bibinfo{author}{\bibfnamefont{E.}~\bibnamefont{Ruiz~Arriola}},
  \bibnamefont{and}
  \bibinfo{author}{\bibfnamefont{K.}~\bibnamefont{Golec-Biernat}},
  \bibinfo{journal}{Phys. Rev.} \textbf{\bibinfo{volume}{D77}},
  \bibinfo{pages}{034023} (\bibinfo{year}{2008}{\natexlab{a}}),
  \eprint{0712.1012}.

\bibitem[{\citenamefont{Frederico et~al.}(2010)\citenamefont{Frederico, Pace,
  Pasquini, and Salme}}]{Frederico:2009pj}
\bibinfo{author}{\bibfnamefont{T.}~\bibnamefont{Frederico}},
  \bibinfo{author}{\bibfnamefont{E.}~\bibnamefont{Pace}},
  \bibinfo{author}{\bibfnamefont{B.}~\bibnamefont{Pasquini}}, \bibnamefont{and}
  \bibinfo{author}{\bibfnamefont{G.}~\bibnamefont{Salme}},
  \bibinfo{journal}{Nucl. Phys. B (Proc. Supp.)}
  \textbf{\bibinfo{volume}{199}}, \bibinfo{pages}{264} (\bibinfo{year}{2010}),
  \eprint{0911.1736}.

\bibitem[{\citenamefont{Frederico et~al.}(2009)\citenamefont{Frederico, Pace,
  Pasquini, and Salme}}]{Frederico:2009fk}
\bibinfo{author}{\bibfnamefont{T.}~\bibnamefont{Frederico}},
  \bibinfo{author}{\bibfnamefont{E.}~\bibnamefont{Pace}},
  \bibinfo{author}{\bibfnamefont{B.}~\bibnamefont{Pasquini}}, \bibnamefont{and}
  \bibinfo{author}{\bibfnamefont{G.}~\bibnamefont{Salme}},
  \bibinfo{journal}{Phys. Rev.} \textbf{\bibinfo{volume}{D80}},
  \bibinfo{pages}{054021} (\bibinfo{year}{2009}), \eprint{0907.5566}.

\bibitem[{\citenamefont{Polyakov}(1999)}]{Polyakov:1998ze}
\bibinfo{author}{\bibfnamefont{M.~V.} \bibnamefont{Polyakov}},
  \bibinfo{journal}{Nucl. Phys.} \textbf{\bibinfo{volume}{B555}},
  \bibinfo{pages}{231} (\bibinfo{year}{1999}), \eprint{hep-ph/9809483}.

\bibitem[{\citenamefont{Esaibegian and Tamarian}(1990)}]{Esaibegian:1989uj}
\bibinfo{author}{\bibfnamefont{S.~V.} \bibnamefont{Esaibegian}}
  \bibnamefont{and} \bibinfo{author}{\bibfnamefont{S.~N.}
  \bibnamefont{Tamarian}}, \bibinfo{journal}{Sov. J. Nucl. Phys.}
  \textbf{\bibinfo{volume}{51}}, \bibinfo{pages}{310} (\bibinfo{year}{1990}).

\bibitem[{\citenamefont{Dorokhov}(1996)}]{Dorokhov:1991nj}
\bibinfo{author}{\bibfnamefont{A.~E.} \bibnamefont{Dorokhov}},
  \bibinfo{journal}{Nuovo Cim.} \textbf{\bibinfo{volume}{A109}},
  \bibinfo{pages}{391} (\bibinfo{year}{1996}).

\bibitem[{\citenamefont{Petrov et~al.}(1999)\citenamefont{Petrov, Polyakov,
  Ruskov, Weiss, and Goeke}}]{Petrov:1998kg}
\bibinfo{author}{\bibfnamefont{V.~Y.} \bibnamefont{Petrov}},
  \bibinfo{author}{\bibfnamefont{M.~V.} \bibnamefont{Polyakov}},
  \bibinfo{author}{\bibfnamefont{R.}~\bibnamefont{Ruskov}},
  \bibinfo{author}{\bibfnamefont{C.}~\bibnamefont{Weiss}}, \bibnamefont{and}
  \bibinfo{author}{\bibfnamefont{K.}~\bibnamefont{Goeke}},
  \bibinfo{journal}{Phys. Rev.} \textbf{\bibinfo{volume}{D59}},
  \bibinfo{pages}{114018} (\bibinfo{year}{1999}), \eprint{hep-ph/9807229}.

\bibitem[{\citenamefont{Anikin et~al.}(2000{\natexlab{b}})\citenamefont{Anikin,
  Dorokhov, and Tomio}}]{Anikin:1999cx}
\bibinfo{author}{\bibfnamefont{I.~V.} \bibnamefont{Anikin}},
  \bibinfo{author}{\bibfnamefont{A.~E.} \bibnamefont{Dorokhov}},
  \bibnamefont{and} \bibinfo{author}{\bibfnamefont{L.}~\bibnamefont{Tomio}},
  \bibinfo{journal}{Phys. Lett.} \textbf{\bibinfo{volume}{B475}},
  \bibinfo{pages}{361} (\bibinfo{year}{2000}{\natexlab{b}}),
  \eprint{hep-ph/9909368}.

\bibitem[{\citenamefont{Praszalowicz and
  Rostworowski}(2001)}]{Praszalowicz:2001wy}
\bibinfo{author}{\bibfnamefont{M.}~\bibnamefont{Praszalowicz}}
  \bibnamefont{and}
  \bibinfo{author}{\bibfnamefont{A.}~\bibnamefont{Rostworowski}},
  \bibinfo{journal}{Phys. Rev.} \textbf{\bibinfo{volume}{D64}},
  \bibinfo{pages}{074003} (\bibinfo{year}{2001}), \eprint{hep-ph/0105188}.

\bibitem[{\citenamefont{Dorokhov}(2003)}]{Dorokhov:2002iu}
\bibinfo{author}{\bibfnamefont{A.~E.} \bibnamefont{Dorokhov}},
  \bibinfo{journal}{JETP Lett.} \textbf{\bibinfo{volume}{77}},
  \bibinfo{pages}{63} (\bibinfo{year}{2003}), \eprint{hep-ph/0212156}.

\bibitem[{\citenamefont{Ruiz~Arriola and
  Broniowski}(2002)}]{RuizArriola:2002bp}
\bibinfo{author}{\bibfnamefont{E.}~\bibnamefont{Ruiz~Arriola}}
  \bibnamefont{and}
  \bibinfo{author}{\bibfnamefont{W.}~\bibnamefont{Broniowski}},
  \bibinfo{journal}{Phys. Rev.} \textbf{\bibinfo{volume}{D66}},
  \bibinfo{pages}{094016} (\bibinfo{year}{2002}), \eprint{hep-ph/0207266}.

\bibitem[{\citenamefont{Ruiz~Arriola}(2002)}]{RuizArriola:2002wr}
\bibinfo{author}{\bibfnamefont{E.}~\bibnamefont{Ruiz~Arriola}},
  \bibinfo{journal}{Acta Phys. Polon.} \textbf{\bibinfo{volume}{B33}},
  \bibinfo{pages}{4443} (\bibinfo{year}{2002}), \eprint{hep-ph/0210007}.

\bibitem[{\citenamefont{Broniowski and Ruiz~Arriola}(2008)}]{Broniowski:2008hx}
\bibinfo{author}{\bibfnamefont{W.}~\bibnamefont{Broniowski}} \bibnamefont{and}
  \bibinfo{author}{\bibfnamefont{E.}~\bibnamefont{Ruiz~Arriola}},
  \bibinfo{journal}{Phys. Rev.} \textbf{\bibinfo{volume}{D78}},
  \bibinfo{pages}{094011} (\bibinfo{year}{2008}), \eprint{0809.1744}.

\bibitem[{\citenamefont{Pire and
  Szymanowski}(2005{\natexlab{a}})}]{Pire:2004ie}
\bibinfo{author}{\bibfnamefont{B.}~\bibnamefont{Pire}} \bibnamefont{and}
  \bibinfo{author}{\bibfnamefont{L.}~\bibnamefont{Szymanowski}},
  \bibinfo{journal}{Phys. Rev.} \textbf{\bibinfo{volume}{D71}},
  \bibinfo{pages}{111501} (\bibinfo{year}{2005}{\natexlab{a}}),
  \eprint{hep-ph/0411387}.

\bibitem[{\citenamefont{Pire and
  Szymanowski}(2005{\natexlab{b}})}]{Pire:2005ax}
\bibinfo{author}{\bibfnamefont{B.}~\bibnamefont{Pire}} \bibnamefont{and}
  \bibinfo{author}{\bibfnamefont{L.}~\bibnamefont{Szymanowski}},
  \bibinfo{journal}{Phys. Lett.} \textbf{\bibinfo{volume}{B622}},
  \bibinfo{pages}{83} (\bibinfo{year}{2005}{\natexlab{b}}),
  \eprint{hep-ph/0504255}.

\bibitem[{\citenamefont{Lansberg et~al.}(2006)\citenamefont{Lansberg, Pire, and
  Szymanowski}}]{Lansberg:2006fv}
\bibinfo{author}{\bibfnamefont{J.~P.} \bibnamefont{Lansberg}},
  \bibinfo{author}{\bibfnamefont{B.}~\bibnamefont{Pire}}, \bibnamefont{and}
  \bibinfo{author}{\bibfnamefont{L.}~\bibnamefont{Szymanowski}},
  \bibinfo{journal}{Phys. Rev.} \textbf{\bibinfo{volume}{D73}},
  \bibinfo{pages}{074014} (\bibinfo{year}{2006}), \eprint{hep-ph/0602195}.

\bibitem[{\citenamefont{Lansberg et~al.}(2007)\citenamefont{Lansberg, Pire, and
  Szymanowski}}]{Lansberg:2007bu}
\bibinfo{author}{\bibfnamefont{J.~P.} \bibnamefont{Lansberg}},
  \bibinfo{author}{\bibfnamefont{B.}~\bibnamefont{Pire}}, \bibnamefont{and}
  \bibinfo{author}{\bibfnamefont{L.}~\bibnamefont{Szymanowski}}, in
  \emph{\bibinfo{booktitle}{proc. of Exclusive Reactions at High Momentum
  Transfer, Jefferson Lab, 2007}}, edited by
  \bibinfo{editor}{\bibfnamefont{A.}~\bibnamefont{Radyushkin}}
  \bibnamefont{and} \bibinfo{editor}{\bibfnamefont{P.}~\bibnamefont{Stoler}}
  (\bibinfo{publisher}{World Scientific eBooks}, \bibinfo{year}{2007}),
  \eprint{0709.2567}.

\bibitem[{\citenamefont{Tiburzi}(2005)}]{Tiburzi:2005nj}
\bibinfo{author}{\bibfnamefont{B.~C.} \bibnamefont{Tiburzi}},
  \bibinfo{journal}{Phys. Rev.} \textbf{\bibinfo{volume}{D72}},
  \bibinfo{pages}{094001} (\bibinfo{year}{2005}), \eprint{hep-ph/0508112}.

\bibitem[{\citenamefont{Broniowski and Ruiz~Arriola}(2007)}]{Broniowski:2007fs}
\bibinfo{author}{\bibfnamefont{W.}~\bibnamefont{Broniowski}} \bibnamefont{and}
  \bibinfo{author}{\bibfnamefont{E.}~\bibnamefont{Ruiz~Arriola}},
  \bibinfo{journal}{Phys. Lett.} \textbf{\bibinfo{volume}{B649}},
  \bibinfo{pages}{49} (\bibinfo{year}{2007}), \eprint{hep-ph/0701243}.

\bibitem[{\citenamefont{Courtoy and Noguera}(2007)}]{Courtoy:2007vy}
\bibinfo{author}{\bibfnamefont{A.}~\bibnamefont{Courtoy}} \bibnamefont{and}
  \bibinfo{author}{\bibfnamefont{S.}~\bibnamefont{Noguera}},
  \bibinfo{journal}{Phys. Rev.} \textbf{\bibinfo{volume}{D76}},
  \bibinfo{pages}{094026} (\bibinfo{year}{2007}), \eprint{0707.3366}.

\bibitem[{\citenamefont{Courtoy and Noguera}(2008)}]{Courtoy:2008af}
\bibinfo{author}{\bibfnamefont{A.}~\bibnamefont{Courtoy}} \bibnamefont{and}
  \bibinfo{author}{\bibfnamefont{S.}~\bibnamefont{Noguera}},
  \bibinfo{journal}{Prog. Part. Nucl. Phys.} \textbf{\bibinfo{volume}{61}},
  \bibinfo{pages}{170} (\bibinfo{year}{2008}), \eprint{0803.3524}.

\bibitem[{\citenamefont{Kotko and Praszalowicz}(2009)}]{Kotko:2008gy}
\bibinfo{author}{\bibfnamefont{P.}~\bibnamefont{Kotko}} \bibnamefont{and}
  \bibinfo{author}{\bibfnamefont{M.}~\bibnamefont{Praszalowicz}},
  \bibinfo{journal}{Acta Phys. Polon.} \textbf{\bibinfo{volume}{B40}},
  \bibinfo{pages}{123} (\bibinfo{year}{2009}), \eprint{0803.2847}.

\bibitem[{\citenamefont{Sutton et~al.}(1992)\citenamefont{Sutton, Martin,
  Roberts, and Stirling}}]{Sutton:1991ay}
\bibinfo{author}{\bibfnamefont{P.~J.} \bibnamefont{Sutton}},
  \bibinfo{author}{\bibfnamefont{A.~D.} \bibnamefont{Martin}},
  \bibinfo{author}{\bibfnamefont{R.~G.} \bibnamefont{Roberts}},
  \bibnamefont{and} \bibinfo{author}{\bibfnamefont{W.~J.}
  \bibnamefont{Stirling}}, \bibinfo{journal}{Phys. Rev.}
  \textbf{\bibinfo{volume}{D45}}, \bibinfo{pages}{2349} (\bibinfo{year}{1992}).

\bibitem[{\citenamefont{Gluck et~al.}(1999)\citenamefont{Gluck, Reya, and
  Schienbein}}]{Gluck:1999xe}
\bibinfo{author}{\bibfnamefont{M.}~\bibnamefont{Gluck}},
  \bibinfo{author}{\bibfnamefont{E.}~\bibnamefont{Reya}}, \bibnamefont{and}
  \bibinfo{author}{\bibfnamefont{I.}~\bibnamefont{Schienbein}},
  \bibinfo{journal}{Eur. Phys. J.} \textbf{\bibinfo{volume}{C10}},
  \bibinfo{pages}{313} (\bibinfo{year}{1999}), \eprint{hep-ph/9903288}.

\bibitem[{\citenamefont{Best et~al.}(1997)}]{Best:1997qp}
\bibinfo{author}{\bibfnamefont{C.}~\bibnamefont{Best}} \bibnamefont{et~al.},
  \bibinfo{journal}{Phys. Rev.} \textbf{\bibinfo{volume}{D56}},
  \bibinfo{pages}{2743} (\bibinfo{year}{1997}), \eprint{hep-lat/9703014}.

\bibitem[{\citenamefont{Diakonov and Petrov}(1986)}]{Diakonov:1985eg}
\bibinfo{author}{\bibfnamefont{D.}~\bibnamefont{Diakonov}} \bibnamefont{and}
  \bibinfo{author}{\bibfnamefont{V.~Y.} \bibnamefont{Petrov}},
  \bibinfo{journal}{Nucl. Phys.} \textbf{\bibinfo{volume}{B272}},
  \bibinfo{pages}{457} (\bibinfo{year}{1986}).

\bibitem[{\citenamefont{Holdom et~al.}(1990)\citenamefont{Holdom, Terning, and
  Verbeek}}]{Holdom:1990iq}
\bibinfo{author}{\bibfnamefont{B.}~\bibnamefont{Holdom}},
  \bibinfo{author}{\bibfnamefont{J.}~\bibnamefont{Terning}}, \bibnamefont{and}
  \bibinfo{author}{\bibfnamefont{K.}~\bibnamefont{Verbeek}},
  \bibinfo{journal}{Phys. Lett.} \textbf{\bibinfo{volume}{B245}},
  \bibinfo{pages}{612} (\bibinfo{year}{1990}).

\bibitem[{\citenamefont{Maris et~al.}(1998)\citenamefont{Maris, Roberts, and
  Tandy}}]{Maris:1997hd}
\bibinfo{author}{\bibfnamefont{P.}~\bibnamefont{Maris}},
  \bibinfo{author}{\bibfnamefont{C.~D.} \bibnamefont{Roberts}},
  \bibnamefont{and} \bibinfo{author}{\bibfnamefont{P.~C.} \bibnamefont{Tandy}},
  \bibinfo{journal}{Phys. Lett.} \textbf{\bibinfo{volume}{B420}},
  \bibinfo{pages}{267} (\bibinfo{year}{1998}), \eprint{nucl-th/9707003}.

\bibitem[{\citenamefont{Bogolyubov and Shirkov}(Wiley, New York,
  1980)}]{Bogolyubov:1980}
\bibinfo{author}{\bibfnamefont{N.}~\bibnamefont{Bogolyubov}} \bibnamefont{and}
  \bibinfo{author}{\bibfnamefont{D.}~\bibnamefont{Shirkov}}
  (\bibinfo{year}{Wiley, New York, 1980}).

\bibitem[{\citenamefont{Zavialov}(Kluwer Academic, Dordrecht,
  1990)}]{Zavialov:1990}
\bibinfo{author}{\bibfnamefont{O.}~\bibnamefont{Zavialov}}
  (\bibinfo{year}{Kluwer Academic, Dordrecht, 1990}).

\bibitem[{\citenamefont{Radyushkin}(1997)}]{Radyushkin:1997ki}
\bibinfo{author}{\bibfnamefont{A.~V.} \bibnamefont{Radyushkin}},
  \bibinfo{journal}{Phys. Rev.} \textbf{\bibinfo{volume}{D56}},
  \bibinfo{pages}{5524} (\bibinfo{year}{1997}), \eprint{hep-ph/9704207}.

\bibitem[{\citenamefont{Dorokhov}(2010{\natexlab{a}})}]{Dorokhov:2010bz}
\bibinfo{author}{\bibfnamefont{A.~E.} \bibnamefont{Dorokhov}}
  (\bibinfo{year}{2010}{\natexlab{a}}), \eprint{1003.4693}.

\bibitem[{\citenamefont{Dorokhov}(2010{\natexlab{b}})}]{Dorokhov:2010zzb}
\bibinfo{author}{\bibfnamefont{A.~E.} \bibnamefont{Dorokhov}},
  \bibinfo{journal}{JETP Lett.} \textbf{\bibinfo{volume}{92}},
  \bibinfo{pages}{707} (\bibinfo{year}{2010}{\natexlab{b}}).

\bibitem[{\citenamefont{Radyushkin}(2011)}]{Radyushkin:2011dh}
\bibinfo{author}{\bibfnamefont{A.~V.} \bibnamefont{Radyushkin}},
  \bibinfo{journal}{Phys. Rev.} \textbf{\bibinfo{volume}{D83}},
  \bibinfo{pages}{076006} (\bibinfo{year}{2011}), \eprint{1101.2165}.

\bibitem[{\citenamefont{Broniowski
  et~al.}(2008{\natexlab{b}})\citenamefont{Broniowski, Ruiz~Arriola, and
  Golec-Biernat}}]{BAG}
\bibinfo{author}{\bibfnamefont{W.}~\bibnamefont{Broniowski}},
  \bibinfo{author}{\bibfnamefont{E.}~\bibnamefont{Ruiz~Arriola}},
  \bibnamefont{and}
  \bibinfo{author}{\bibfnamefont{K.}~\bibnamefont{Golec-Biernat}},
  \bibinfo{journal}{Phys. Rev.} \textbf{\bibinfo{volume}{D77}},
  \bibinfo{pages}{034023} (\bibinfo{year}{2008}{\natexlab{b}}),
  \eprint{0712.1012}.

\bibitem[{\citenamefont{Perevalova et~al.}(2011)\citenamefont{Perevalova,
  Polyakov, Vall, and Vladimirov}}]{Perevalova:2011qi}
\bibinfo{author}{\bibfnamefont{I.~A.} \bibnamefont{Perevalova}},
  \bibinfo{author}{\bibfnamefont{M.~V.} \bibnamefont{Polyakov}},
  \bibinfo{author}{\bibfnamefont{A.~N.} \bibnamefont{Vall}}, \bibnamefont{and}
  \bibinfo{author}{\bibfnamefont{A.~A.} \bibnamefont{Vladimirov}}
  (\bibinfo{year}{2011}), \eprint{1105.4990}.

\bibitem[{\citenamefont{Gribov}(1973)}]{Gribov:1973jg}
\bibinfo{author}{\bibfnamefont{V.}~\bibnamefont{Gribov}}
  (\bibinfo{year}{1973}), \bibinfo{note}{in *Moscow 1 ITEP school, v.1
  'Elementary particles'*, 65,1973}, \eprint{hep-ph/0006158}.

\bibitem[{\citenamefont{Gasser and Leutwyler}(1984)}]{Gasser:1983yg}
\bibinfo{author}{\bibfnamefont{J.}~\bibnamefont{Gasser}} \bibnamefont{and}
  \bibinfo{author}{\bibfnamefont{H.}~\bibnamefont{Leutwyler}},
  \bibinfo{journal}{Ann. Phys.} \textbf{\bibinfo{volume}{158}},
  \bibinfo{pages}{142} (\bibinfo{year}{1984}).

\bibitem[{\citenamefont{Broniowski}(1999)}]{Broniowski:1999dm}
\bibinfo{author}{\bibfnamefont{W.}~\bibnamefont{Broniowski}}
  (\bibinfo{year}{1999}), \eprint{hep-ph/9911204}.

\bibitem[{\citenamefont{Pagels and Stokar}(1979)}]{Pagels:1979hd}
\bibinfo{author}{\bibfnamefont{H.}~\bibnamefont{Pagels}} \bibnamefont{and}
  \bibinfo{author}{\bibfnamefont{S.}~\bibnamefont{Stokar}},
  \bibinfo{journal}{Phys. Rev.} \textbf{\bibinfo{volume}{D20}},
  \bibinfo{pages}{2947} (\bibinfo{year}{1979}).

\bibitem[{\citenamefont{Ruiz~Arriola}(1991)}]{RuizArriola:1991gc}
\bibinfo{author}{\bibfnamefont{E.}~\bibnamefont{Ruiz~Arriola}},
  \bibinfo{journal}{Phys. Lett.} \textbf{\bibinfo{volume}{B253}},
  \bibinfo{pages}{430} (\bibinfo{year}{1991}).

\bibitem[{\citenamefont{Schuren et~al.}(1992)\citenamefont{Schuren,
  Ruiz~Arriola, and Goeke}}]{Schuren:1991sc}
\bibinfo{author}{\bibfnamefont{C.}~\bibnamefont{Schuren}},
  \bibinfo{author}{\bibfnamefont{E.}~\bibnamefont{Ruiz~Arriola}},
  \bibnamefont{and} \bibinfo{author}{\bibfnamefont{K.}~\bibnamefont{Goeke}},
  \bibinfo{journal}{Nucl. Phys.} \textbf{\bibinfo{volume}{A547}},
  \bibinfo{pages}{612} (\bibinfo{year}{1992}).

\bibitem[{\citenamefont{Christov et~al.}(1996)}]{Christov:1995vm}
\bibinfo{author}{\bibfnamefont{C.~V.} \bibnamefont{Christov}}
  \bibnamefont{et~al.}, \bibinfo{journal}{Prog. Part. Nucl. Phys.}
  \textbf{\bibinfo{volume}{37}}, \bibinfo{pages}{91} (\bibinfo{year}{1996}),
  \eprint{hep-ph/9604441}.

\bibitem[{\citenamefont{Gerasimov}(1979)}]{Gerasimov:1978cp}
\bibinfo{author}{\bibfnamefont{S.~B.} \bibnamefont{Gerasimov}},
  \bibinfo{journal}{Yad. Fiz.} \textbf{\bibinfo{volume}{29}},
  \bibinfo{pages}{513} (\bibinfo{year}{1979}).

\bibitem[{\citenamefont{Pivovarov}(2003)}]{Pivovarov:2001mw}
\bibinfo{author}{\bibfnamefont{A.~A.} \bibnamefont{Pivovarov}},
  \bibinfo{journal}{Phys. Atom. Nucl.} \textbf{\bibinfo{volume}{66}},
  \bibinfo{pages}{902} (\bibinfo{year}{2003}), \eprint{hep-ph/0110248}.

\bibitem[{\citenamefont{Boughezal and Melnikov}(2011)}]{Boughezal:2011vw}
\bibinfo{author}{\bibfnamefont{R.}~\bibnamefont{Boughezal}} \bibnamefont{and}
  \bibinfo{author}{\bibfnamefont{K.}~\bibnamefont{Melnikov}}
  (\bibinfo{year}{2011}), \eprint{1104.4510}.

\bibitem[{\citenamefont{Milton et~al.}(2001)\citenamefont{Milton, Solovtsov,
  and Solovtsova}}]{Milton:2001mq}
\bibinfo{author}{\bibfnamefont{K.~A.} \bibnamefont{Milton}},
  \bibinfo{author}{\bibfnamefont{I.~L.} \bibnamefont{Solovtsov}},
  \bibnamefont{and} \bibinfo{author}{\bibfnamefont{O.~P.}
  \bibnamefont{Solovtsova}}, \bibinfo{journal}{Phys. Rev.}
  \textbf{\bibinfo{volume}{D64}}, \bibinfo{pages}{016005}
  (\bibinfo{year}{2001}), \eprint{hep-ph/0102254}.

\bibitem[{\citenamefont{Dorokhov}(2004)}]{Dorokhov:2004ze}
\bibinfo{author}{\bibfnamefont{A.~E.} \bibnamefont{Dorokhov}},
  \bibinfo{journal}{Phys. Rev.} \textbf{\bibinfo{volume}{D70}},
  \bibinfo{pages}{094011} (\bibinfo{year}{2004}), \eprint{hep-ph/0405153}.

\bibitem[{\citenamefont{Broniowski and Ruiz~Arriola}(2009)}]{Broniowski:evol}
\bibinfo{author}{\bibfnamefont{W.}~\bibnamefont{Broniowski}} \bibnamefont{and}
  \bibinfo{author}{\bibfnamefont{E.}~\bibnamefont{Ruiz~Arriola}},
  \bibinfo{journal}{Phys. Rev.} \textbf{\bibinfo{volume}{D79}},
  \bibinfo{pages}{057501} (\bibinfo{year}{2009}), \eprint{0901.3336}.

\bibitem[{\citenamefont{Kivel and
  Mankiewicz}(1999{\natexlab{a}})}]{Kivel:1999sk}
\bibinfo{author}{\bibfnamefont{N.}~\bibnamefont{Kivel}} \bibnamefont{and}
  \bibinfo{author}{\bibfnamefont{L.}~\bibnamefont{Mankiewicz}},
  \bibinfo{journal}{Phys. Lett.} \textbf{\bibinfo{volume}{B458}},
  \bibinfo{pages}{338} (\bibinfo{year}{1999}{\natexlab{a}}),
  \eprint{hep-ph/9905342}.

\bibitem[{\citenamefont{Kivel and
  Mankiewicz}(1999{\natexlab{b}})}]{Kivel:1999wa}
\bibinfo{author}{\bibfnamefont{N.}~\bibnamefont{Kivel}} \bibnamefont{and}
  \bibinfo{author}{\bibfnamefont{L.}~\bibnamefont{Mankiewicz}},
  \bibinfo{journal}{Nucl. Phys.} \textbf{\bibinfo{volume}{B557}},
  \bibinfo{pages}{271} (\bibinfo{year}{1999}{\natexlab{b}}),
  \eprint{hep-ph/9903531}.

\bibitem[{\citenamefont{Manashov et~al.}(2005)\citenamefont{Manashov, Kirch,
  and Schafer}}]{Manashov:2005xp}
\bibinfo{author}{\bibfnamefont{A.}~\bibnamefont{Manashov}},
  \bibinfo{author}{\bibfnamefont{M.}~\bibnamefont{Kirch}}, \bibnamefont{and}
  \bibinfo{author}{\bibfnamefont{A.}~\bibnamefont{Schafer}},
  \bibinfo{journal}{Phys. Rev. Lett.} \textbf{\bibinfo{volume}{95}},
  \bibinfo{pages}{012002} (\bibinfo{year}{2005}), \eprint{hep-ph/0503109}.

\bibitem[{\citenamefont{Kirch et~al.}(2005)\citenamefont{Kirch, Manashov, and
  Schafer}}]{Kirch:2005tt}
\bibinfo{author}{\bibfnamefont{M.}~\bibnamefont{Kirch}},
  \bibinfo{author}{\bibfnamefont{A.}~\bibnamefont{Manashov}}, \bibnamefont{and}
  \bibinfo{author}{\bibfnamefont{A.}~\bibnamefont{Schafer}},
  \bibinfo{journal}{Phys. Rev.} \textbf{\bibinfo{volume}{D72}},
  \bibinfo{pages}{114006} (\bibinfo{year}{2005}), \eprint{hep-ph/0509330}.

\bibitem[{\citenamefont{Brommel et~al.}(2006)}]{Brommel:2005ee}
\bibinfo{author}{\bibfnamefont{D.}~\bibnamefont{Brommel}} \bibnamefont{et~al.},
  \bibinfo{journal}{PoS} \textbf{\bibinfo{volume}{LAT2005}},
  \bibinfo{pages}{360} (\bibinfo{year}{2006}), \eprint{hep-lat/0509133}.

\end{thebibliography}
\end{document}